\documentclass[acmsmall,nonacm]{acmart}

\usepackage{subfig}
\usepackage{algorithm}
\usepackage{algpseudocode}

\AtBeginDocument{%
  \providecommand\BibTeX{{%
    \normalfont B\kern-0.5em{\scshape i\kern-0.25em b}\kern-0.8em\TeX}}}

\begin{document}

\title{Dynamic Path-Decomposed Tries}

\author{Shunsuke Kanda}
\email{shunsuke.kanda@riken.jp}
\affiliation{%
  \institution{RIKEN Center for Advanced Intelligence Project}
  \country{Japan}
}

\author{Dominik K{\"o}ppl}
\email{dominik.koeppl@inf.kyushu-u.ac.jp}
\affiliation{%
  \institution{Kyushu University}
  \country{Japan}
}
\affiliation{%
  \institution{Japan Society for Promotion of Science}
}

\author{Yasuo Tabei}
\email{yasuo.tabei@riken.jp}
\affiliation{%
  \institution{RIKEN Center for Advanced Intelligence Project}
  \country{Japan}
}

\author{Kazuhiro Morita}
\email{kam@is.tokushima-u.ac.jp}
\affiliation{%
  \institution{Tokushima University}
  \country{Japan}
}

\author{Masao Fuketa}
\email{fuketa@is.tokushima-u.ac.jp}
\affiliation{%
  \institution{Tokushima University}
  \country{Japan}
}

\renewcommand{\shortauthors}{Kanda et al.}

\begin{abstract}
A keyword dictionary is an associative array whose keys are strings.
Recent applications handling massive keyword dictionaries in main memory have a need for a space-efficient implementation.
When limited to static applications, there are a number of highly-compressed keyword dictionaries based on the advancements of practical succinct data structures.
However, as most succinct data structures are only efficient in the static case,  it is still difficult to implement a keyword dictionary that is \emph{space efficient} and \emph{dynamic}.
In this article, we propose such a keyword dictionary.
Our main idea is to embrace the path decomposition technique, which was proposed for constructing cache-friendly tries.
To store the path-decomposed trie in small memory, we design data structures based on recent compact hash trie representations. %
Experiments on real-world datasets reveal that our dynamic keyword dictionary needs up to 68\% less space than the existing smallest ones, while achieving a relevant space-time tradeoff.
\end{abstract}

\maketitle

\newcommand{\Str}[1]{\texttt{#1}}
\newcommand{\Ceil}[1]{\lceil{#1}\rceil}
\newcommand{\Floor}[1]{\lfloor{#1}\rfloor}
\newcommand{\Tuple}[1]{({#1})}
\newcommand{\Order}{\mathcal{O}}
\newcommand{\Polylog}{\mathrm{polylog}}

\newcommand{\Trie}{\mathcal{T}}
\newcommand{\PDT}{\mathcal{T}^{c}}
\newcommand{\DynPDT}{\mathcal{T}^{c}}
\newcommand{\True}{\top}
\newcommand{\False}{\bot}

\newcommand{\Lookup}{\textsf{lookup}}
\newcommand{\Insert}{\textsf{insert}}
\newcommand{\Delete}{\textsf{delete}}

\newcommand{\Dict}{\mathcal{S}}
\newcommand{\DictB}{\mathcal{D}}
\newcommand{\AlphA}{\mathcal{A}}
\newcommand{\AlphB}{\mathcal{B}}
\newcommand{\Univ}{\mathcal{X}}

\newcommand{\AddChild}{\textsf{addchild}}
\newcommand{\GetChild}{\textsf{getchild}}
\newcommand{\GetParent}{\textsf{getparent}}
\newcommand{\GetEdge}{\textsf{getedge}}

\newcommand{\fref}[1]{Figure \ref{#1}}
\newcommand{\tref}[1]{Table \ref{#1}}
\newcommand{\trefs}[2]{Tables \ref{#1} and \ref{#2}}
\newcommand{\sref}[1]{Section \ref{#1}}
\newcommand{\srefs}[2]{Sections \ref{#1} and \ref{#2}}
\newcommand{\gref}[1]{Algorithm \ref{#1}}
\newcommand{\pref}[1]{Property \ref{#1}}
\newcommand{\aref}[1]{Appendix \ref{#1}}

\newcommand{\PBT}{\textsf{PBT}}
\newcommand{\CBT}{\textsf{CBT}}
\newcommand{\PHT}{\textsf{PFKT}}
\newcommand{\CHT}{\textsf{CFKT}}

\newcommand{\PLM}{\textsf{PLM}}
\newcommand{\CLM}{\textsf{SLM}}

\newcommand{\PPBT}{\textsf{PDT-PB}}
\newcommand{\PCBT}{\textsf{PDT-SB}}
\newcommand{\CCBT}{\textsf{PDT-CB}}
\newcommand{\PPHT}{\textsf{PDT-PFK}}
\newcommand{\PCHT}{\textsf{PDT-SFK}}
\newcommand{\CCHT}{\textsf{PDT-CFK}}

\newcommand{\ArHash}{\textsf{ArrayHash}}
\newcommand{\Judy}{\textsf{Judy}}
\newcommand{\HAT}{\textsf{HAT}}
\newcommand{\Cedar}{\textsf{Cedar}}
\newcommand{\CedarR}{\textsf{Cedar-R}}
\newcommand{\CedarP}{\textsf{Cedar-P}}
\newcommand{\STL}{\textsf{STLHash}}
\newcommand{\GDense}{\textsf{GoogleDense}}
\newcommand{\GSparse}{\textsf{Sparsepp}}
\newcommand{\Hopscotch}{\textsf{Hopscotch}}
\newcommand{\Robin}{\textsf{Robin}}
\newcommand{\ART}{\textsf{ART}}
\newcommand{\PCTB}{\textsf{PCT-Bit}}
\newcommand{\PCTH}{\textsf{PCT-Hash}}
\newcommand{\ZFT}{\textsf{ZFT}}
\newcommand{\CTrie}{\textsf{CTrie++}}

\newcommand{\Geo}{\textsf{GeoNames}}
\newcommand{\Wiki}{\textsf{Wiki}}
\newcommand{\AOL}{\textsf{AOL}}
\newcommand{\DNA}{\textsf{DNA}}
\newcommand{\UK}{\textsf{UK}}
\newcommand{\WebBase}{\textsf{WebBase}}
\newcommand{\LUBMS}{\textsf{LUBMS}}
\newcommand{\LUBML}{\textsf{LUBML}}

\newcommand{\Size}{\textsf{Size}}
\newcommand{\MinLen}{\textsf{MinLen}}
\newcommand{\MaxLen}{\textsf{MaxLen}}
\newcommand{\AveLen}{\textsf{AveLen}}
\newcommand{\Steps}{\textsf{Steps}}
\newcommand{\Space}{\textsf{Space}}
\newcommand{\Time}{\textsf{Time}}
\newcommand{\AveNLL}{\textsf{AveNLL}}
\newcommand{\InsertTime}{\textsf{Insert}}
\newcommand{\LookupTime}{\textsf{Lookup}}

\newcommand{\AveHeight}{\textsf{AveHeight}}
\newcommand{\AveHeightLB}{\textsf{AveHeightLB}}
\newcommand{\AveHeightUB}{\textsf{AveHeightUB}}

\newcommand{\PlotWidth}{18em}
\newcommand{\FigScale}{0.6}

\renewcommand{\algorithmicrequire}{\textbf{Input:}}
\renewcommand{\algorithmicensure}{\textbf{Output:}}

\algnewcommand{\IfThen}[2]{%
  \State \algorithmicif\ #1\ \algorithmicthen\ #2}

\theoremstyle{definition}
\newtheorem{property}{Property}

\section{Introduction}
\label{sect:intro}

An associative array is called a \emph{keyword dictionary} if its keys are strings.
In this article, we study the problem to maintain a keyword dictionary in main memory efficiently.
When storing words extracted from text collections written in natural or computer languages,
the size of a keyword dictionary~$A$ is not of major concern.
This is because, after carefully polishing the extracted strings with natural language processing tools like stemmers, the size of~$A$  grows sublinearly as $\Order(N^\beta)$ for some $\beta \approx 0.5$ over a text of $N$ words due to Heaps' Law \cite{books:heaps1978information,books:baeza2011modern}.
However, as reported in \cite{martinez2016practical}, some natural language applications such as web search engines and machine translation systems need to handle large datasets that are not under Heaps' Law.
Also, other recent applications as in Semantic Web graphs and in bioinformatics handle massive string databases with keyword dictionaries \cite{martinez2016practical,mavlyutov2015comparison}.
Although common implementations like hash tables are fast, their memory consumption is a severe drawback in such scenarios.
Here, a \emph{space-efficient} implementation of the keyword dictionary is important.
In this paper, we focus on the \emph{practical} side of this problem.

In the static setting, omitting the insertion and deletion of keywords, a number of compressed keyword dictionaries have been developed for a decade, some of which we highlight in the following.
We start with Mart{\'\i}nez-Prieto et al. \cite{martinez2016practical}, who proposed and evaluated a number of compressed keyword dictionaries based on techniques like hashing, front-coding, full-text indexes, and tries.
They demonstrated that their implementations use up to 5\% space of the original dataset size, while also supporting searches of prefixes and substrings of the keywords.
Subsequently, Grossi and Ottaviano \cite{grossi2014fast} proposed a cache-friendly keyword dictionary through path decomposition of tries.
Arz and Fischer \cite{arz2018lempel} adapted the LZ78 compression to devise a keyword dictionary.
Finally, Kanda et al. \cite{kanda2017compressed} proposed a keyword dictionary based on a compressed double-array trie.
As we can see from these representations, space-efficient static keyword dictionaries have been well studied because of the advancements of practical (yet static) succinct data structures collected in well maintained libraries such as SDSL \cite{gog2014theory} and Succinct \cite{grossi2013design}.

Under the dynamic setting, however, only a few space-efficient keyword dictionaries have been realized, probably due to the implementation difficulty.
Although HAT-trie \cite{askitis2010engineering} and Judy \cite{manual:judy10min} are representative space-efficient dynamic implementations as demonstrated in previous experiments\footnote{Such as \url{http://www.tkl.iis.u-tokyo.ac.jp/~ynaga/cedar/\#perf} and \url{https://github.com/Tessil/hat-trie/blob/master/README.md\#benchmark}.}, they still waste memory by maintaining many pointers.
The Cedar trie \cite{yoshinaga2014self} is a space-efficient implementation embracing heavily 32-bit pointers to address memory, and therefore cannot be applied to massive datasets.
Its implementation makes it hard to switch to 64-bit pointers, but we expect that doing so will increase its space consumption considerably.
Although several practical dynamic succinct data structures \cite{prezza2017framework,poyias2017compact,poyias2018mbonsai} have been recently developed, modern dynamic keyword dictionaries are heavily based on pointers, consuming a large fraction of the entire space requirement.
Nonetheless, there are some applications that need dynamic keyword dictionaries for massive datasets such as search engines \cite{software:groonga,busch2012earlybird}, RDF stores \cite{mavlyutov2015comparison}, or Web crawler \cite{ueda2013parallel}.
Consequently, realizing a practical space-efficient dynamic keyword dictionaries is an important open challenge.

\subsection{Space-Efficient Dynamic Tries}
Common keyword dictionary implementations represent the keywords in a trie, supporting the retrieval of keywords with trie navigation operations.
In this subsection, we summarize space-efficient dynamic tries.

\paragraph{Theoretical Discussion}
We consider a dynamic trie with $t$ nodes over an alphabet of size $\sigma$.
Arroyuelo et al. \cite{arroyuelo2016succinct} introduced succinct representations that require almost optimal $2t + t \log \sigma + o(t \log \sigma)$ bits of space, while supporting insertion and deletion of a leaf in $\Order(1)$ amortized time if $\sigma = \Order(\Polylog(t))$ and in $\Order(\log \sigma / \log \log \sigma)$ amortized time otherwise.\footnote{Throughout this paper, the base of the logarithm is 2, whenever not explicitly indicated.}
Jansson et al. \cite{jansson2015linked} presented a dynamic trie representation that uses $\Order(t \log \sigma)$ bits of space, while supporting insertion and deletion of a leaf in $\Order(\log \log t)$ expected amortized time.

\paragraph{Hash Tries}
On the practical side, Poyias et al. \cite{poyias2018mbonsai} proposed the \emph{m-Bonsai} trie, a practical dynamic compact trie representation.
It is a variant of the Bonsai trie \cite{darragh1993bonsai} that
represents the trie nodes as entries in a compact hash table.
It takes $\Order(t\log\sigma)$ bits of space, while supporting update and some traversal operations in $\Order(1)$ expected time.
Fischer and K{\"{o}}ppl \cite{fischer2017practical} presented and evaluated a number of dynamic tries for LZ78 \cite{ziv1978compression} and LZW \cite{welch1984technique} factorization.
They also proposed an efficient hash-based trie representation in a similar way to m-Bonsai, which is referred to as \emph{FK-hash}.\footnote{The representation is referred to as \textsf{hash} or \textsf{cht} in their paper \cite{fischer2017practical}. To avoid confusion, we name it FK-hash by using the initial letters of the proposers, Fischer and K{\"{o}}ppl.}
Although FK-hash uses $\Order(t\log\sigma + t\log t)$ bits of space, its update algorithm is simple and practically fast.
However, we are not aware of any space-efficient approach using them as keyword dictionaries.

\paragraph{Compacted Tries}
Another line of research focuses on limiting the space of the trie in relation to the number of keywords.
Suppose that we want to maintain a set of $n$ strings with a total length of $N$ on a machine, where $\beta = \log_{\sigma} N$ characters fit into a single machine word $w$.
In this setting, Belazzougui et al. \cite{belazzougui2010dynamic} proposed the (dynamic) z-fast trie, which takes $N \log \sigma + \Order(n \log N)$ bits of space and supports retrieval, insertion and deletion of a string $S$ in $\Order(|S|/\beta + \log |S| + \log\log \sigma)$
expected time.
Takagi et al. \cite{takagi2016packed} proposed the packed compact trie, which takes $N \log \sigma + \Order(nw)$ bits of space and supports the same operations in $\Order(|S|/\beta + \log\log N)$ expected time.
Recently, Tsuruta et al. \cite{tsuruta2020ctrie} developed a hybrid data structure of the z-fast trie and the packed compact trie, which also takes $N \log \sigma + \Order(nw)$ bits of space, but improves each of these operations to run in $\Order(|S|/\beta + \log \beta)$ expected time.

\subsection{Our Contribution}

We propose a novel space-efficient dynamic keyword dictionary, called the \emph{dynamic path-decomposed trie} (abbreviated as \emph{DynPDT}).
DynPDT is based on a trie formed by \textit{path decomposition} \cite{ferragina2008searching}.
The path decomposition is a trie transformation technique, which
was proposed for constructing cache-friendly trie dictionaries.
It was up to now utilized only in static applications \cite{grossi2014fast,hsu2013space}. Here, we adapt this technique for the dynamic construction of DynPDT,
which gives DynPDT two main advantages over other known keyword dictionaries.
\begin{enumerate}
\item The first is that the data structure is cache efficient because of the path decomposition.
During the retrieval of a keyword, most parts of the keyword can be scanned in a cache-friendly manner without node-to-node traversals based on random accesses.
\item The second is that the path decomposition allows us to plug in any dynamic trie representation for the path-decomposed trie topology. For this job, we choose the hash-based trie representations m-Bonsai and FK-hash as these are fast and memory efficient in the setting when all trie nodes have to be represented explicitly (which is the case for the nodes of the path-decomposed trie).
\end{enumerate}
Based on these advantages, DynPDT becomes a fast and space-efficient dynamic keyword dictionary.

From experiments using massive real-world datasets, we demonstrate that DynPDT is more space efficient compared to existing keyword dictionaries while achieving a relevant space-time tradeoff.
For example, to construct a keyword dictionary from a large URI dataset of 13.8 GiB, DynPDT needs only 2.5 GiB of working space, while a HAT-trie and a Judy trie need 9.5 GiB and 7.8 GiB, respectively.
The time performance is competitive in many cases thanks to the path decomposition.
The source code of our implementation is available at \url{https://github.com/kampersanda/poplar-trie}.

\subsection{Paper Structure}

In \sref{sect:pre}, we introduce the keyword dictionary, and review the trie data structure and the path decomposition in our preliminaries.
We introduce our new data structure DynPDT in \sref{sect:dynpdt}.
Subsequently, we present our DynPDT representations based on m-Bonsai and FK-hash in \srefs{sect:bonsai}{sect:fkhash}, respectively.
In \sref{sect:exp}, we provide our experimental results.
Finally, we conclude the paper in \sref{sect:con}.\footnote{A preliminary version of this work appeared in our conference paper \cite{kanda2017practical} and the first author's Ph.D. thesis \cite{kanda2018space}. This paper contains the significant differences as follows: (1) a fast variant of m-Bonsai was incorporated in \sref{sect:bonsai:plain}; (2) an efficient implementation of the bijective hash function in m-Bonsai was incorporated in \sref{sect:bonsai:compact}; (3) a growing algorithm of m-Bonsai was presented in \sref{sect:bonsai:grow}; (4) FK-hash was also considered in addition to m-Bonsai in \sref{sect:fkhash}; (5) the experimental results in \sref{sect:exp} and all descriptions were significantly enhanced.}
\section{Preliminaries}
\label{sect:pre}

A \emph{string} is a (finite) sequence of characters over a finite alphabet.
Our strings always start at position 0.
Given a string $S$ of length $n$, $S[i,j)$ denotes the \emph{substring} $S[i],S[i+1],\ldots,S[j-1]$ for $0 \leq i \leq j \leq n$.
Particularly, $S[0,j)$ is a \emph{prefix} of $S$ and $S[i,n)$ is a \emph{suffix} of $S$.
Let $|S| := n$ denote the length of $S$.
The same notation is also applied to \emph{arrays}.
The cardinality of a set $A$ is denoted by $|A|$.

Our model of computation is the transdichotomous word RAM model of word size $w = \Theta(\log N)$, where $N$ is the total length of all keywords of a given problem, i.e., the size of the problem.
We can read and process $\Order(w)$ bits in constant time.

\subsection{Keyword Dictionary}\label{sec:keywordict}

A \emph{keyword} is a string over an alphabet $\AlphA$ that is terminated with a special character $\Str{\$} \not\in \AlphA$ at its end.
In a \emph{prefix-free} set of strings, no string is a prefix of another string.
A set of keywords is always prefix-free due to the character \Str{\$}.
A \emph{keyword dictionary} is a dynamic associative array that maps a dynamic set of $n$ keywords $\Dict = \{K_1, K_2, ..., K_{n}\} \subset \AlphA^{*}$ to values $x_1,x_2,\ldots,x_n$, where $x_i$ belongs to a finite set $\Univ$.
It supports the retrieval, the insertion, and the deletion of keywords while maintaining the \emph{key-value mapping}.
In detail, it supports the following operations:
 
\begin{itemize}
    \item $\Lookup(K)$ returns the value associated with the keyword $K$ if $K \in \Dict$ or $\False$ otherwise.
    \item $\Insert(K,x)$ inserts the keyword $K$ in $\Dict$, i.e., $\Dict \gets \Dict \cup \{K\}$, and associates the value $x$ with $K$.
    \item $\Delete(K)$ removes the keyword $K$ from $\Dict$, i.e., $\Dict \gets \Dict \setminus \{K\}$.
\end{itemize}

\subsection{Tries}

A trie \cite{books:knuth1998art,fredkin1960trie} is a rooted labeled tree $\Trie_\Dict$ representing a set of keywords $\Dict$.
Each edge in $\Trie_\Dict$ is labeled by a character.
All outgoing edges of a node are labeled with a distinct character.
The label $c$ of the edge $(u,v)$ between a node $v$ and its parent $u$ is called the \emph{branching character} of $v$. The parent $u$ and branching character $c$ unique determines $v$. 
Each keyword $K \in \Dict$ is represented by exactly one path from the root to a leaf $u$, i.e., the keyword $K$ can be extracted by concatenating the edge labels on the path from the root to $u$.
Since $\Dict$ is prefix-free ($\Str{\$}$ is a unique delimiter of each keyword), there is a 1-to-1 correlation between leaves and keywords.

Given a keyword $K$ of length $m$, $\Trie_\Dict$ retrieves $K$ by traversing nodes from the root to a leaf while matching the characters of $K$ with the edge labels of the traversed path.
In representations storing all trie nodes explicitly, we visit $m$ nodes during this traversal.
However, this traversal suffers poor locality of reference since it needs to access pointers usually addressing non-consecutive memory.
In practice, this cache inefficiency is a critical bottleneck especially for long strings such as URLs.
Grossi and Ottaviano \cite{grossi2014fast} successfully solved this problem through \emph{path decomposition} \cite{ferragina2008searching} in practice (but, in static settings).

\subsection{Path Decomposition}
\label{sect:pre:pd}

The \emph{path decomposition} \cite{ferragina2008searching} of a trie $\Trie_\Dict$ is a recursive procedure that first chooses an arbitrary root-to-leaf path $\pi$ in $\Trie_\Dict$, then compactifies the path $\pi$ to a single node, and subsequently repeats the procedure in each subtrie hanging off the path $\pi$.
As a result, $\Trie_\Dict$ is partitioned into a set of $n$ node-to-leaf paths because there are $n$ leaves in $\Trie_\Dict$.
This decomposition produces the \emph{path-decomposed} trie $\PDT_\Dict$, which is composed of $n$ compactified nodes.

\begin{figure}[tb]
\centering
\includegraphics[scale=\FigScale]{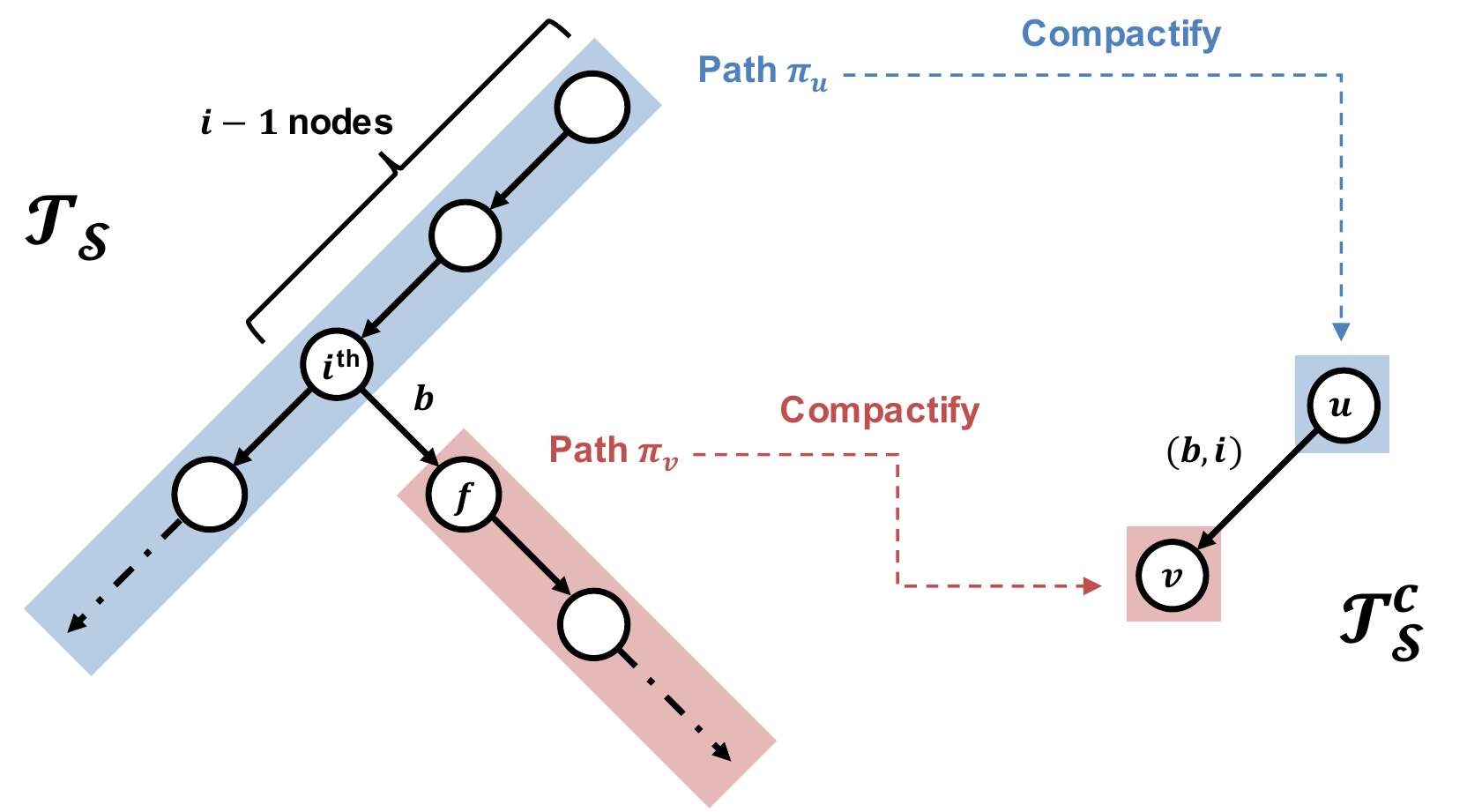}
\label{fig:pd}
\caption{Illustration of the path decomposition of the path $\pi_v$ whose first node~$f$ is a child of the $i$-th node on the path $\pi_u$ represented by node $u$ in $\PDT_\Dict$.
Since the branching character of~$f$ is~$b$, $u$ and $v$ are connected with an edge with label $\Tuple{b,i}$.}
\label{fig:pd}
\end{figure}

For explaining the properties of $\PDT_\Dict$, we call the concatenation of the labels of all edges of a node-to-leaf path $\pi$ in $\Trie_\Dict$ the \emph{path string} of $\pi$. 
The path strings of the compactified paths of $\Trie_\Dict$ are the node labels of $\PDT_\Dict$. In detail,
each node $u$ in $\PDT_\Dict$ is associated with a node-to-leaf path $\pi$ of $\Trie_\Dict$ and is labeled by the path string of $\pi$, denoted by $L_u \in \AlphA^*$.
Each edge in $\PDT_\Dict$ is labeled by a pair consisting of a branching character and an integer, which are defined as follows (see also \fref{fig:pd}):
Take a node~$u$ in $\PDT_\Dict$ and one of its children $v$.
Suppose that $u$ and $v$ are associated with the paths $\pi_u$ and $\pi_v$ in $\Trie_\Dict$, respectively, such that $L_u$ and $L_v$ are the path labels of $\pi_u$ and $\pi_v$.
The edge $\Tuple{u,v}$ has the label $\Tuple{b,i}$ if, in $\Trie_\Dict$, the first node on the path~$\pi_v$ is the node
\begin{itemize}
\item whose branching character is~$b$, and
\item whose parent is the $i$-th node\footnote{Throughout this paper, we start counting from zero.} visited on the path~$\pi_u$. 
\end{itemize}
The edge labels of $\PDT_\Dict$ are characters drawn from the alphabet $\AlphB := \AlphA \times \{0,1,\ldots,\Lambda - 1 \}$, where $\Lambda$ is the longest length of all node labels.

\begin{figure}[tb]
\centering
\subfloat[Trie $\Trie_\Dict$]{
    \includegraphics[scale=\FigScale]{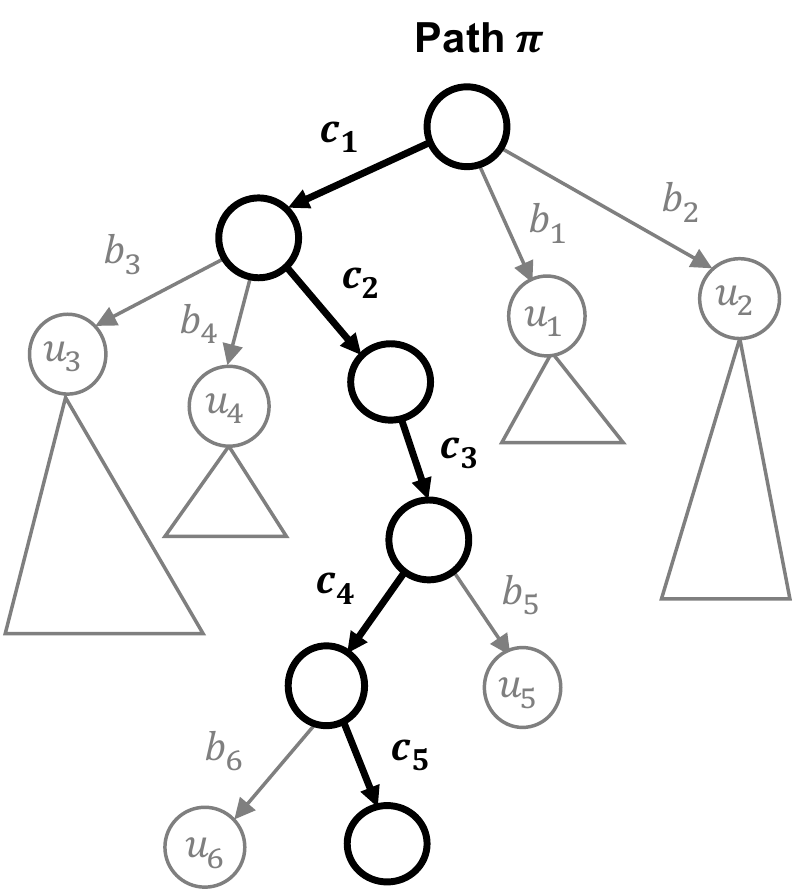}
    \label{fig:trie}
}
\subfloat[Path-decomposed trie $\PDT_\Dict$]{
    \includegraphics[scale=\FigScale]{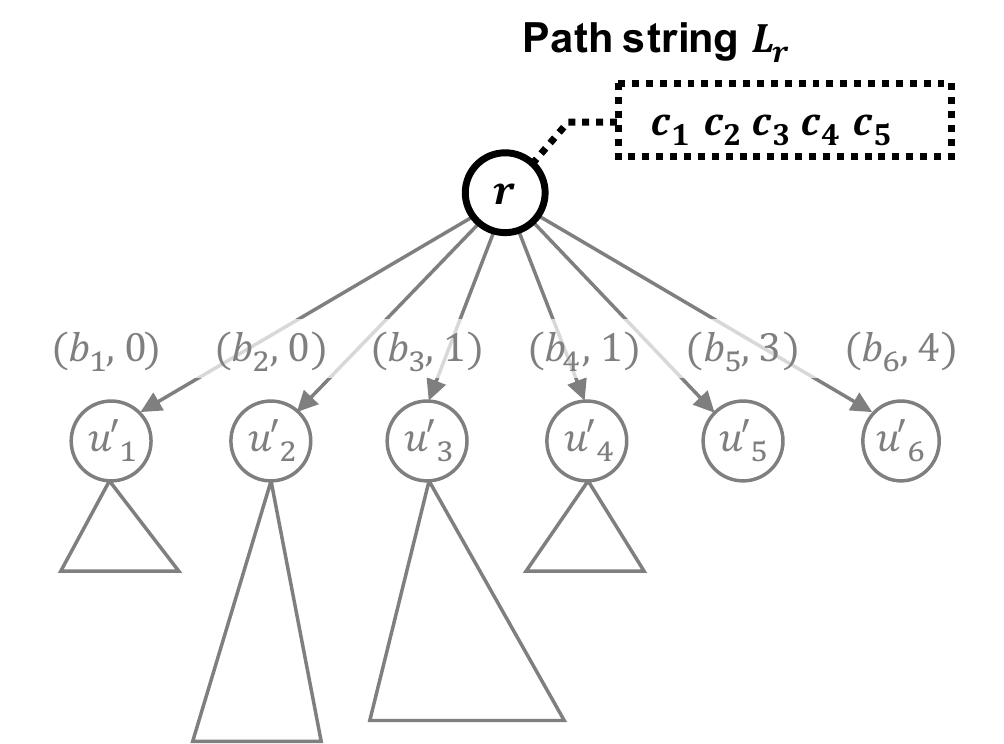}
    \label{fig:pdt}
}
\caption{Illustration of the first transformation of the path decomposition.}
\label{fig:pathdec}
\end{figure}

\begin{example}[Path-Decomposed Trie]
\fref{fig:pathdec} illustrates a root-to-leaf path $\pi$ in $\Trie_\Dict$ and the corresponding root $r$ in $\PDT_\Dict$ after compactifying $\pi$ to $r$.
The root $r$ is labeled by the path string of $\pi$, which is $L_r = c_1 c_2 c_3 c_4 c_5$.
The branching character of $u'_5$ in $\PDT_\Dict$ is $\Tuple{b_5,3}$ because $u_5$ in $\Trie_\Dict$ is the child of the third node on the path $\pi$ with branching character $b_5$.
Also for the subtries rooted at the nodes $u_1, u_2, \ldots, u_6$ in $\Trie_\Dict$, the decomposition is recursively applied to produce the children of the root in $\PDT_\Dict$.
\end{example}

Given a keyword $K$, the retrieval on $\Trie_\Dict$ can be simulated with a traversal of $\PDT_\Dict$ starting at its root:
Let $u$ denote the currently visited node in $\PDT_\Dict$.
On visiting $u$, we compare the path string $L_u$ with the characters of $K$.
If we find a mismatch at $L_u[i]$ with $b := K[i] \neq L_u[i]$, 
we descend to the child with branching character $\Tuple{b,i}$
and drop the first $i+1$ characters of $K$.

When storing the characters of each path string $L_u$ in consecutive memory locations, the number of random accesses involved in the retrieval on $\PDT_\Dict$ is bounded by $\Order(h)$, where $h$ is the height of $\PDT_\Dict$.
The following property regarding the height is satisfied by construction.

\begin{property}
\label{prop:pd}
The height of $\PDT_\Dict$ cannot be larger than that of $\Trie_\Dict$.
\end{property}

\paragraph{Centroid Path Decomposition}
A way to improve this height bound in the static case is the \emph{centroid} path decomposition \cite{ferragina2008searching}.
Given an inner node $u$ in $\Trie_\Dict$, the \emph{heavy} child of $u$ is the child whose subtrie has the most leaves (ties are broken arbitrarily).
Given a node $u$, the \emph{centroid path} is the path from $u$ to a leaf obtained by descending only to heavy children.
The centroid path decomposition yields the following property by always choosing centroid paths in the decomposition.

\begin{property}[\cite{ferragina2008searching}]
\label{prop:cent}
Through the centroid path decomposition, the height of $\PDT_\Dict$ is bounded by $\Order(\log n)$.
\end{property}

\paragraph{Key-Value Mapping}

We can implement the key-value mapping through $\PDT_\Dict$ because there is a 1-to-1 correlation between nodes in $\PDT_\Dict$ and keywords in $\Dict$.
A simple approach is to store the associated values in an array $A$ such that $A[u]$ stores the value associated with node $u$. If we assign each of the $n$ nodes in $\PDT_\Dict$ a unique id from the range $[0,n)$, then
$A$ has no vacant entry (i.e. $|A| = n$).
Another approach is to embed the value of~$K_i$ at the end of $L_u$, where the node $u$ corresponds to the keyword $K_i$.
This approach can be used without considering the assignment of node ids.
In our experiments, we used the latter approach.

\section{Dynamic Path-Decomposed Trie}
\label{sect:dynpdt}

Although the centroid path decomposition gives a logarithmic upper bound on the height of $\PDT_\Dict$ (cf.\ \sref{prop:cent}), it can be adapted only in static settings because we have to know the complete topology of $\Trie_\Dict$ \emph{a priori} to determine the centroid paths.
As a matter of fact, previous data structures embracing the path decomposition \cite{grossi2014fast,hsu2013space,ferragina2008searching} consider only static applications.

In this section, we present the \emph{incremental} path decomposition, which is a novel procedure to construct a \emph{dynamic} path-decomposed trie, which we call DynPDT in the following.
Our procedure incrementally chooses\footnote{We actually do not construct $\Trie_\Dict$, but represent it with the DynPDT $\PDT_\Dict$} a node-to-leaf path in $\Trie_\Dict$ and directly updates the DynPDT $\PDT_\Dict$ on inserting a new keyword of $\Dict$.
This incrementally chosen path is not a centroid path in general.
Thus, the incremental path decomposition does not necessarily satisfy \pref{prop:cent} but always satisfies \pref{prop:pd}.

In this section, we drop the technical detail of storing the values to ease 
the explanation of DynPDT, for which we omit the second argument in the insert operation
$\Insert(K)$ of a new keyword~$K$.

\subsection{Incremental Path Decomposition}
\label{sect:dynpdt:ipd}

In the following, we simulate a dynamic trie $\Trie_\Dict$ by DynPDT $\DynPDT_\Dict$.
Suppose that $\Trie_\Dict$ is non-empty.
On inserting a new keyword $K \not\in \Dict$ into $\Trie_\Dict$, we proceed as follows:
\begin{enumerate}
\item First traverse $\Trie_{\Dict}$ from the root by matching characters of $K$ until reaching the deepest node $u$ whose string label $X$ is a prefix of $K$.
\item Decompose $K$ into $K = XbY$ for $b \in \AlphA$ and $Y \in \AlphA^{*}$, which is possible since $K \not\in \Dict$ and $K[|K|-1]= \Str{\$}$.
\item Finally, insert a new child $v$ of $u$ with branching character $b$ and append, from node $v$, new nodes corresponding to the suffix $Y$.
\end{enumerate}

In other words, the task of $\Insert(K)$ on $\Trie_{\Dict}$ is to create a new node-to-leaf path $\pi$ representing the suffix $Y$.
We call that path $\pi$ the \emph{incremental path} of the keyword $K$.
We simulate $\Insert(K)$  by creating a new node in $\DynPDT_\Dict$ whose label is the path label of this incremental path $\pi$:

\def\suf{S}

\begin{itemize}
    \item If $\Dict = \emptyset$, create the root $u_1$ and associate the keyword $K$ with $u_1$ by $L_{u_1} \gets K$. 
    \item Otherwise ($\Dict \neq \emptyset$), retrieve the keyword $K$ from the root $u_1$ in three steps after setting variables $u \gets u_1$ and $\suf \gets K$:
    \begin{enumerate}
        \item Compare $\suf$ with $L_u$. If $\suf = L_u$, terminate because $K$ is already inserted; otherwise, proceed with Step 2.
        \item Find $i$ such that $\suf[0,i) = L_u[0,i)$ and $\suf[i] \neq L_u[i]$ ($i$ exists since $K \not\in \Dict$ and $K[|K|-1] = \texttt{\$}$), and search the child of $u$ with branching character $\Tuple{\suf[i],i}$.
        If found, go back to Step 1 after setting the variable $u$ to this child and $\suf$ to the remaining suffix $\suf[i+1,|\suf|)$; otherwise, proceed with Step 3.
        \item Insert $K$ into $\Dict$ by creating a new child $v$ of $u$ with branching character $\Tuple{\suf[i],i}$, and store the remaining suffix in $v$ by $L_{v} \gets \suf[i + 1,|\suf|)$.
    \end{enumerate}
\end{itemize}

\begin{figure}[tb]
\centering
\subfloat[$\Insert(\Str{technology\$})$]{
    \includegraphics[scale=\FigScale]{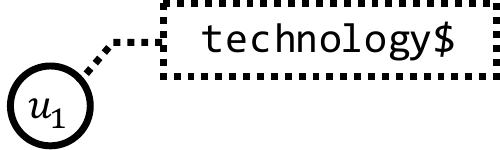}
    \label{fig:ipd-a}
}
\subfloat[$\Insert(\Str{technics\$})$]{
    \includegraphics[scale=\FigScale]{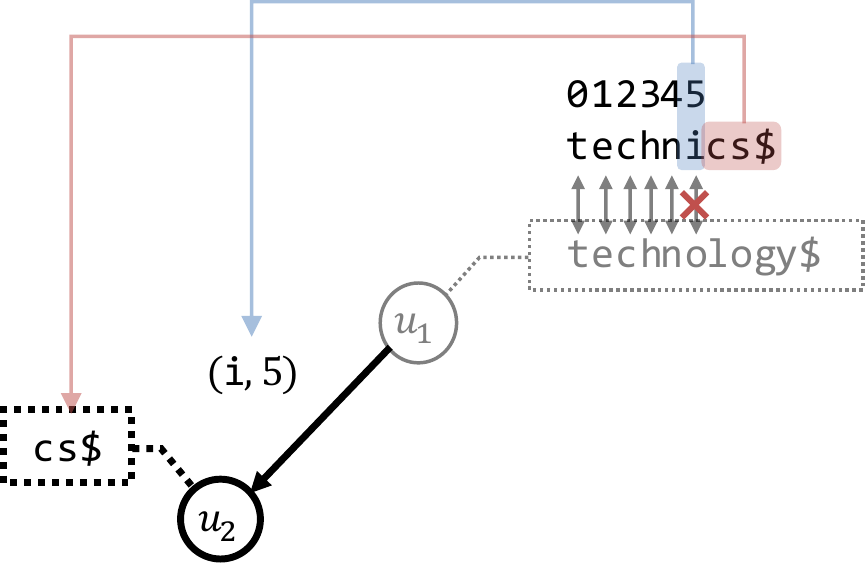}
    \label{fig:ipd-b}
}\\
\subfloat[$\Insert(\Str{technique\$})$]{
    \includegraphics[scale=\FigScale]{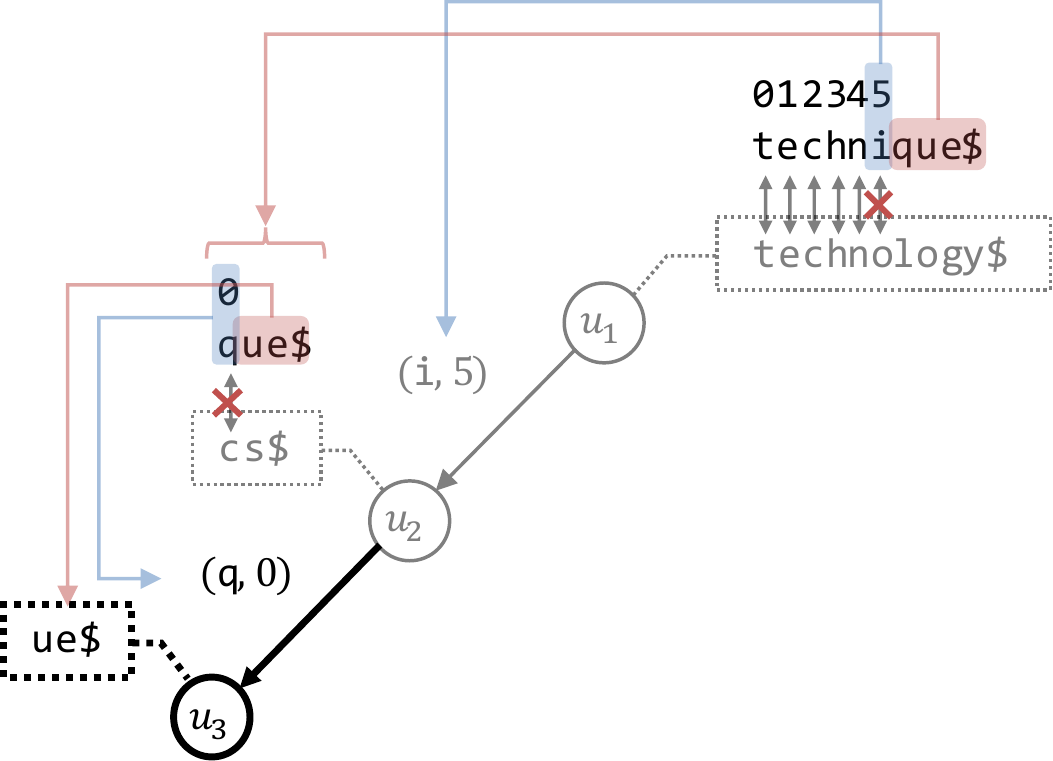}
    \label{fig:ipd-c}
}
\subfloat[$\Insert(\Str{technically\$})$]{
    \includegraphics[scale=\FigScale]{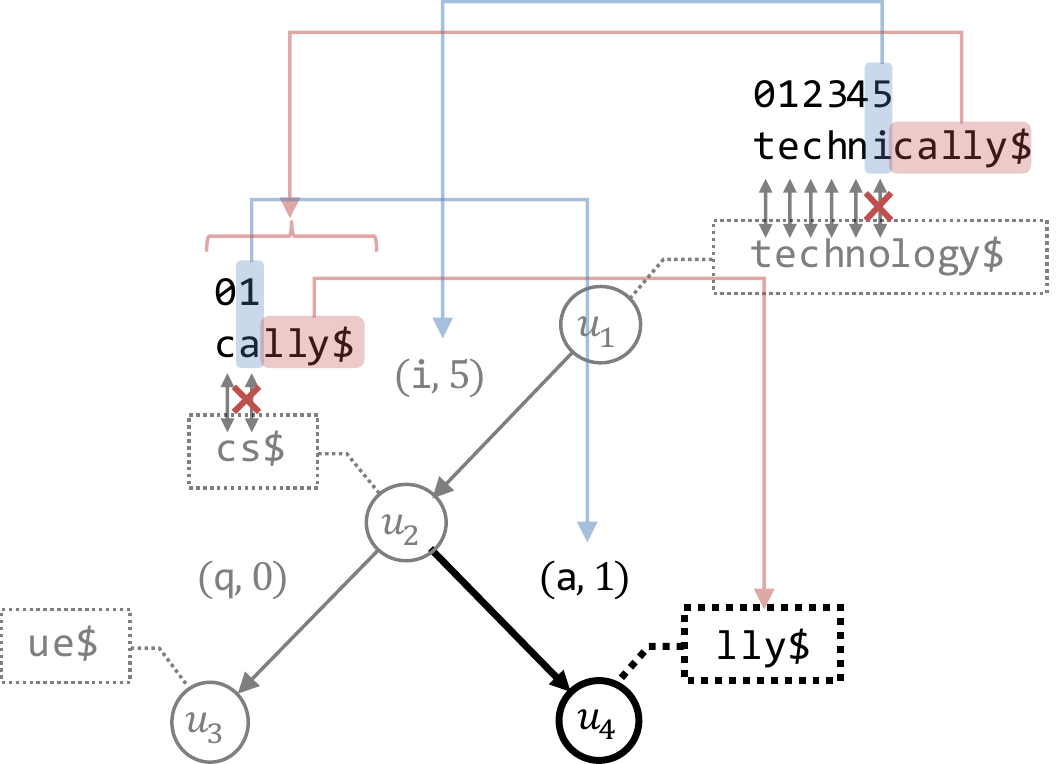}
    \label{fig:ipd-d}
}
\caption{Process of incremental path decomposition for keywords \Str{technology\$}, \Str{technics\$}, \Str{technique\$} and \Str{technically\$} in this order.}
\label{fig:ipd}
\end{figure}

\begin{example}[Construction]\label{ex:construction}

\fref{fig:ipd} illustrates the construction process of DynPDT $\DynPDT_\Dict$ when inserting the keywords $K_1 = \Str{technology\$}$, $K_2 = \Str{technics\$}$, $K_3 = \Str{technique\$}$, and $K_4 = \Str{technically\$}$ in this order, where the $i$-th created node is denoted by $u_i$.
The process begins with an empty trie $\DynPDT_\Dict$.

\begin{enumerate}
\item[(a)] In the first insertion $\Insert(K_1)$, we create the root $u_1$ and associate $K_1$ with $L_{u_1}$,
that is, $L_{u_1}$ becomes $\Str{technology\$}$.
The resulting $\DynPDT_\Dict$ for $\Dict = \{K_1\}$ is shown in \fref{fig:ipd-a}.

\item[(b)] In the second insertion $\Insert(K_2)$, we define a string variable $\suf$ initially set to $\suf \gets K_2$.
We try to retrieve $K_2$ in $\DynPDT_\Dict$ by comparing $\suf$ with $L_{u_1}$, but fail as there is a mismatching character \Str{i} at position 5 with $\suf[0,5) = L_{u_1}[0,5) = \Str{techn}$ and $\suf[5] = \Str{i} \neq \Str{o} = L_{u_1}[5]$.
Based on this mismatch result, we search the child of $u_1$ with branching character $\Tuple{\Str{i},5}$.
However, since there is no such child, we add a new child $u_2$ to $u_1$ with branching character $\Tuple{\Str{i},5}$ and associate the remaining suffix $\suf[6,|\suf|) = \Str{cs\$}$ with $L_{u_2}$.
The resulting $\DynPDT_\Dict$ for $\Dict = \{K_1,K_2\}$ is shown in \fref{fig:ipd-b}.

\item[(c)] In the third insertion $\Insert(K_3)$, we initially set the string variable $\suf$ to $\suf \gets K_3$ and then compare $\suf$ with $L_{u_1}$ in the same manner as the second insertion.
Since $\suf[0,5) = L_{u_1}[0,5) = \Str{techn}$ and $\suf[5] = \Str{i} \neq \Str{o} = L_{u_1}[5]$, we descend to child $u_2$ with branching character $\Tuple{\Str{i},5}$.
After updating $\suf \gets \suf[6,|\suf|) = \Str{que\$}$, we subsequently compare $\suf$ with $L_{u_2}$ to obtain the mismatch character \Str{q} at position 0 with $\suf[0] = \Str{q} \neq \Str{c} = L_{u}[0]$.
We search the child with branching character $\Tuple{\Str{q},0}$, but there is no such child; thus, we create the child $u_3$ and set $L_{u_3}$ to be the remaining suffix $\suf[1,|\suf|) = \Str{ue\$}$.
The resulting $\DynPDT_\Dict$ for $\Dict = \{K_1,K_2,K_3\}$ is shown in \fref{fig:ipd-c}.

\item[(d)] The fourth insertion $\Insert(K_4)$ is also conducted in the same manner.
The final trie $\DynPDT_\Dict$ is shown in \fref{fig:ipd-d}.
\end{enumerate}

\end{example}

\subsection{Dictionary Operations}
\label{sect:dynpdt:impl}

It is left to define the operations  $\Lookup$ and $\Delete$ to make DynPDT a keyword dictionary.
Similar to $\Insert$, the operation $\Lookup$ can be performed by traversing $\DynPDT_\Dict$ from the root.
After matching all the characters of $K$, $\Lookup(K)$ returns the  value associated with the last visited node. 
It returns $\False$ on a mismatch.

\begin{example}[Retrieval]
We provide an example for a successful and an unsuccessful search.
Both examples are similar to the construction described in Example \ref{ex:construction}.

\begin{enumerate}
\item We consider $\Lookup(\Str{technically\$})$ for the $\DynPDT_\Dict$ in \fref{fig:ipd-d}.
We define a string variable $\suf$ initially set to $\suf \gets \Str{technically\$}$, and compare $\suf$ with $L_{u_1}$ to retrieve (a part of) the keyword from the root.
Since $\suf[0,5) = L_{u_1}[0,5) = \Str{techn}$ and $\suf[5] = \Str{i} \neq \Str{o} = L_{u_1}[5]$, 
we descend to child $u_2$ with branching character $\Tuple{\Str{i},5}$.
Subsequently, we update $\suf$ to be the remaining suffix as $\suf \gets \suf[6,|\suf|) = \Str{cally\$}$ and descend to child $u_4$ with branching character $\Tuple{\Str{a},1}$ since $\suf[0,1) = L_{u_1}[0,1) = \Str{c}$ and $\suf[1] = \Str{a} \neq \Str{s} = L_{u_2}[1]$.
Finally, we update $\suf \gets \suf[2,|\suf|) = \Str{lly\$}$ and compare $\suf$ with $L_{u_4}$. As both match, we return the value stored in $u_4$.

\item We consider $\Lookup(\Str{technical\$})$ for the $\DynPDT_\Dict$ in \fref{fig:ipd-d}.
In the same manner as in the above case, we reach node $u_4$ with the prefix \Str{technica} and subsequently compare $\suf = \Str{l\$}$ and $L_{u_4}$.
Since $\suf[0,1) = L_{u_4}[0,1) = \Str{c}$ and $\suf[1] = \Str{\$} \neq L_{u_4}[1] = \Str{l}$, we search a child with branching character $\Tuple{\Str{\$},1}$; however, there is no such child.
As a result, $\Lookup(\Str{technical\$})$ returns $\False$.
\end{enumerate}
\end{example}

The operation $\Delete$ can be implemented by introducing deletion flags for each node (i.e., for each keyword), a trick that is also used in hashing with open addressing~\cite[Chapter 6.4, Algorithm L]{books:knuth1998art}.
In other words, $\Delete(K)$ retrieves $K$ and sets the deletion flag for the node corresponding to $K$.
However, this approach additionally needs one bit for each node.
Another approach is to set the value associated with the deleted keyword to $\bot$ as an invalid value.
This approach does not need additional space for the deletion flags.
Although these approaches do not free up space after deletion, the space is reused for keywords inserted subsequently if the new keywords share sufficiently long prefixes with the deleted ones.

\subsection{Fixing the Alphabet}
\label{sect:dynpdt:fix}

In practice, a critical problem of DynPDT is that the domain of the edge labels $\AlphB$ in $\DynPDT_\Dict$ and the longest length of all node labels $\Lambda$ are not constant in general.
We tackle this problem by limiting the size of $\AlphB$. 
To this end, we introduce a new parameter $\lambda$ to forcibly fix the alphabet as $\AlphB = \AlphA \times \{0,1,\ldots,\lambda -1\}$ in advance.
Within this limitation, suppose that we want to create an edge labeled $\Tuple{c,i}$ from node $u$ with $i \geq \lambda$.
As this label is not in $\AlphB$, we create dummy nodes called \emph{step nodes} with a special character $\phi$ by repeating the following procedure until $i$ becomes less than $\lambda$: add a new child $v$ of $u$ with branching character $\phi$ and recursively set $u \gets v$ and $i \gets i - \lambda$.
$L_u$ is the empty string if $u$ is a step node.

\begin{figure}[tb]
\centering
\includegraphics[scale=\FigScale]{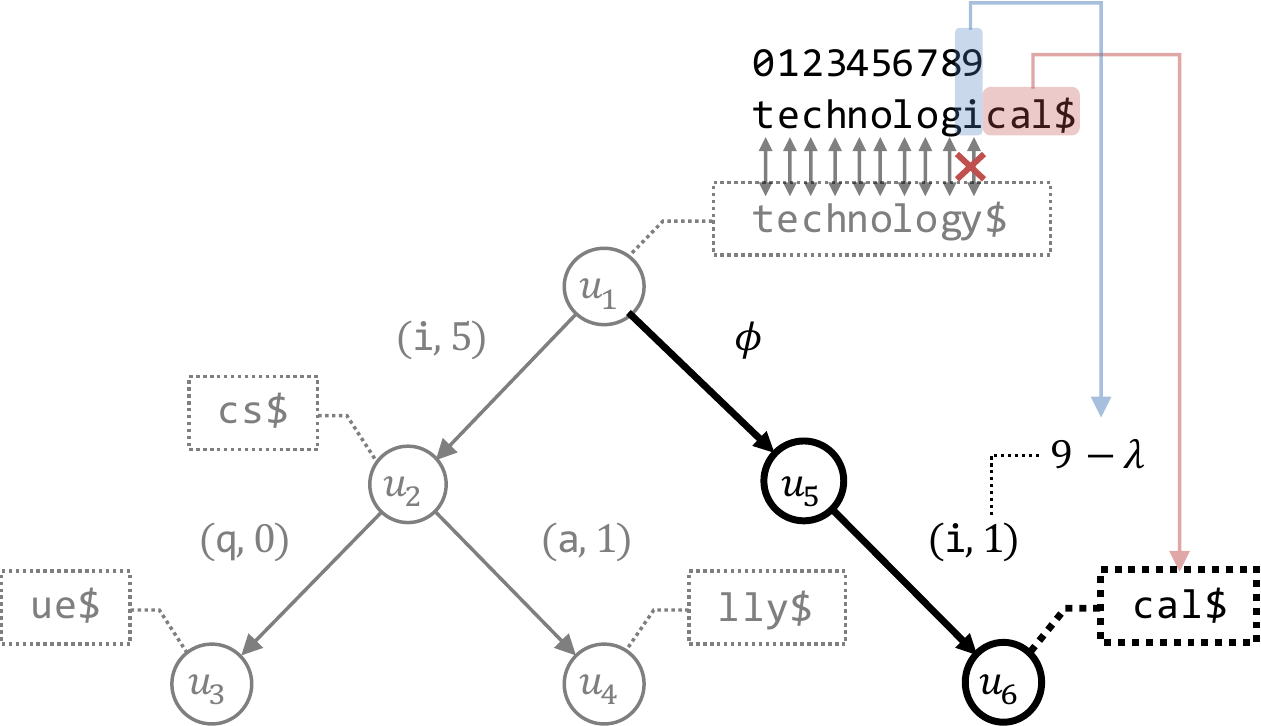}
\label{fig:ipd-e}
\caption{Process of $\Insert(\Str{technological\$})$ when $\lambda = 8$.}
\label{fig:ipd-e}
\end{figure}

\begin{example}[Step Node]

We consider $\Insert(\Str{technological\$})$ for $\DynPDT_\Dict$ in \fref{fig:ipd-d} with $\lambda = 8$.
We set $\suf \gets \Str{technological\$}$ and compare $\suf$ with $L_{u_1}$.
Since $\suf[0,9) = L_{u_1}[0,9) = \Str{technolog}$ and $\suf[9] = \Str{i} \neq \Str{y} = L_{u_1}[9]$, we try to create the edge label $\Tuple{\Str{i},9}$; however, as $i \geq \lambda$, we instead create a step child $u_5$ with branching character $\phi$, descend to this child, and set $i \gets i - \lambda = 1$.
Since $i$ becomes less than $\lambda$, we define a child $u_6$ of the step node $u_5$ with branching character $\Tuple{\Str{i},1}$ and associate the remaining suffix $\suf[10,|\suf|) = \Str{cal\$}$ with $L_{u_6}$.
The resulting DynPDT $\DynPDT_\Dict$ is depicted in \fref{fig:ipd-e}.

\end{example}

This solution creates additional nodes depending on $\lambda$.
When $\lambda$ is too small, many step nodes are created and extra node traversals are involved.
When $\lambda$ is too large, the alphabet size $|\AlphB|$ becomes large and the space usage can increase significantly.
Therefore, it is necessary to determine a suitable $\lambda$.
In \sref{sect:exp}, we empirically determine 32 and 64 to be favorable values for $\lambda$.

\subsection{Representation Scheme}
\label{sect:dynpdt:rep}
To use standard trie techniques, we split up $\DynPDT_\Dict$ into two parts:
\begin{enumerate}
\item a (standard) trie structure $\Trie_\DictB$ for a set of strings $\DictB \subset \AlphB^{*}$ to represent $\DynPDT_\Dict$ with the difference that it assigns a node to a unique id instead of its  node label, and
\item an associative array that maps the ids of the nodes of $\Trie_\DictB$ to their corresponding node labels, called \emph{node label map (NLM)}.
\end{enumerate}
For example, in \fref{fig:ipd-e}, the trie $\Trie_\DictB$ built on the string set $\DictB = \{ \Tuple{\Str{i},5} \Tuple{\Str{q},0}, \Tuple{\Str{i},5} \Tuple{\Str{a},1}, \phi \Tuple{\Str{i},1} \}$ and the NLM stores node labels $L_{u_1},L_{u_2},\ldots,L_{u_6}$ to be accessed by the respective node ids $u_1,u_2,\ldots,u_6$.

\paragraph{Node-Label-Map}
NLM dynamically manages node labels depending on the node ids assigned.
As explained in \sref{sect:intro}, we use the m-Bonsai \cite{poyias2018mbonsai} and FK-hash \cite{fischer2017practical} representations for $\Trie_\DictB$.
Moreover, we design the NLM data structures for m-Bonsai and FK-hash individually, which we respectively present in
\srefs{sect:bonsai}{sect:fkhash}.

\paragraph{Trie Representation $\Trie_\DictB$}
To discuss the representation approaches in the next sections, we define $\Trie_\DictB$ to be a dynamic trie with $n$ nodes whose edge labels are characters drawn from the alphabet $\AlphB$ of size $\sigma = |\AlphA| \cdot \lambda$.
Although the number of nodes~$n$ depends on $\lambda$, we write $n := n(\lambda)$ for simplicity.
$\Trie_\DictB$ supports the following operations:

\begin{itemize}
	\item $\AddChild(u,c)$ adds a new child of $u$ with branching character $c \in \AlphB$ and returns its id.
	\item $\GetChild(u,c)$ returns the id of the child $v$ of~$u$ with branching character $c \in \AlphB$ if $v$ exists, or returns $\False$ otherwise.
\end{itemize}

\paragraph{Motivation for m-Bonsai and FK-hash}
We briefly review some common trie representations and point out their suitability for $\Trie_\DictB$.
The simplest representation is a list trie \cite[Chapter 2.3.2]{askitis2007efficient}, 
which transforms an arbitrary trie to its first-child next-sibling representation.
In this representation, each node of the list trie stores its branching character, a pointer to its first child, and a pointer to its next sibling. 
The list trie represents $\Trie_\DictB$ in $2n\log{n} + n\log{\sigma}$ bits and supports $\AddChild$ and $\GetChild$ in $\Order(\sigma)$ time; however, the operation time becomes problematic if $\sigma = |\AlphA| \cdot \lambda$ is large.
Another representation is a ternary search trie (TST) \cite{bentley1997fast} that reduces the time complexity of the list trie to $\Order(\log\sigma)$; however, the space usage grows to $3n\log{n} + n\log{\sigma}$ bits.
A well-known time- and space-efficient representation is the double array \cite{aoe1989efficient}.
Its space usage is $2n \log n$ bits in the best case, while supporting $\GetChild$ in $\Order(1)$ time; however, a double array for a large alphabet tends to be sparse in practice.
Actually, we are only aware of dynamic double-array implementations handling byte characters (e.g., \cite{yoshinaga2014self,kanda2017rearrangement}).
Judy \cite{manual:judy10min} and ART (adaptive radix tree) \cite{leis2013adaptive} are trie representations that dynamically choose suitable data structures for the trie topology; however, both are also designed for byte characters.
As each trie node is associated with an id, compact tries like the z-fast trie \cite{belazzougui2010dynamic} representing only $O(|\DictB|)$ nodes explicitly become inefficient with this requirement.

Compared to these trie representations, m-Bonsai and FK-hash have better complexities.
m-Bonsai can represent $\Trie_\DictB$ in $cn(\log\sigma + \Order(1))$ bits of expected space for a constant $c > 1$, while supporting $\GetChild$ and $\AddChild$ in $\Order(1)$ expected time \cite{poyias2018mbonsai}. %
Compared to that,
FK-hash needs $cn\log n$ additional bits of expected space, but supports faster insertions in practice.

A straightforward solution to provide the NLM for m-Bonsai and FK-hash is to store the node labels as satellite data in the respective hash table.
However, by doing so, we would waste space for each unoccupied entry in the hash table.
In the following, we present efficient solutions for the NLM tailored to m-Bonsai and FK-hash.

\section{Representation Based on m-Bonsai}
\label{sect:bonsai}

This section presents our approach based on m-Bonsai \cite{poyias2018mbonsai}.
m-Bonsai represents trie nodes as entries in a closed hash table that, spoken informally, compactify the stored keys with {compact hashing} \cite{books:knuth1998art}.

\paragraph{Outline}
We present a plain and a compact form of the $\Trie_\DictB$ representation based on m-Bonsai.
We refer to the former as \PBT{} (Plain m-Bonsai Trie), which is a non-compact variant of m-Bonsai.
\PBT{} can be useful for fast implementation although it has not been considered in any applications yet.
We refer to the latter as \CBT{} (Compact m-Bonsai Trie) as it uses the original m-Bonsai implementation.
We describe \PBT{} and \CBT{} in \srefs{sect:bonsai:plain}{sect:bonsai:compact}, respectively.
In both variants, we maintain a hash table $H$ of size $m$ with the \emph{load factor} $\alpha = n / m \le 1$ to store $n$ nodes.
In \sref{sect:bonsai:grow}, we propose a linear-time growing algorithm based on the approach of Arroyuelo et al. \cite{arroyuelo2017lz78}.
Finally, in \sref{sect:bonsai:nlm}, we propose NLM data structures designed for \PBT{} and \CBT{}.

\subsection{Plain Trie Representation}
\label{sect:bonsai:plain}

\PBT{} uses a hash function $h:\mathbb{N} \rightarrow \mathbb{N}$.
Trie nodes are elements in the hash table.
As their locations in the hash table are fixed unless the hash table is rebuilt, we use these locations as \emph{node ids}.
In other words, the id of a node located at $H[u]$ is $u$.
$\AddChild(u,c)$ is performed as follows.
We first compose the hash key $k = \Tuple{u,c} \in \{0,1,\ldots,m-1\} \times \AlphB$ and then compute its \emph{initial address} $i = h(k) \bmod m$.\footnote{This paper defines $a \bmod b$ as $a - b \cdot \Floor{a/b}$.}
Let $i'$ be the first vacant address from $i$ determined by linear probing.
We create the new child by $H[i'] \gets k$.
That is, the id of the new child becomes $i'$.
$\GetChild$ can be also computed in the same manner.
If $h$ is fully independent and uniformly random, the operations can be performed in $\Order(1)$ expected time.
\PBT{} uses $m\Ceil{\log(m\sigma)}$ bits of space.

\paragraph{Practical Implementation}

The table size $m$ is a power of two in order to quickly compute the modulo operation of $h(k) \bmod m$ by using the bitwise AND operation $h(k) \& (m-1)$ \cite[Section 4.4]{migliore2019masking}.
We set the maximum load factor to $\hat{\alpha} := 0.9$.
If $\alpha$ reaches $\hat{\alpha}$ during an update, we double the size of the hash table by the growing algorithm described in \sref{sect:bonsai:grow}.
We set the initial capacity of the hash table to $m=2^{16}$.
Our hash function~$h$ is a XorShift hash function\footnote{\url{http://xorshift.di.unimi.it/splitmix64.c}.} derived from \cite{steele2014fast}.

\subsection{Compact Trie Representation}
\label{sect:bonsai:compact}

\CBT{} reduces the space usage of \PBT{} with the compact hashing technique \cite{books:knuth1998art}.
Locating nodes on a compact hash table is identical to \PBT{} with the difference that \CBT{} uses a bijective transform $h:\{0,1, \ldots, m\sigma - 1\} \rightarrow \{0,1, \ldots, m\sigma - 1\}$ that maps a key $k$ to its hash value $h(k) \bmod m$ and its quotient $\Floor{h(k)/m}$. Instead of~$k$, the compact hash table stores only its quotient $\Floor{h(k)/m}$ in $H[i']$.
The hash value $h(k)$ can be restored from the initial address $i = h(k) \bmod m$ and the quotient $H[i'] = \Floor{h(k)/m}$, where $i'$ is the first empty slot at or after the initial address~$i$.
The original key $k$ can also be restored from the hash value $h(k)$ since $h$ is bijective.
Therefore, $\AddChild$ and $\GetChild$ can be performed in the same manner as \PBT{} if the corresponding initial address $i$ can be identified from the location $i'$.

The remaining problem is how to identify the corresponding initial address $i$ from $i'$.
Poyias et al. \cite{poyias2018mbonsai} solved this problem by introducing a \emph{displacement array} $D$ such that $D[i']$ keeps the number of probes from $i$ to $i'$, that is, $D[i'] = (i' - i) \bmod m$.
Given a location $i'$, one can compute the corresponding initial address $i$ with $(i' - D[i']) \bmod m$.
Although a value in $D$ is at most $m - 1$, the average value becomes small if $h$ is fully independent and uniformly random and the load factor $\alpha$ is small.
Poyias et al. \cite{poyias2018mbonsai} demonstrated that $D$ can be represented in $\Order(m)$ bits using CDRW (Compact Dynamic ReWritable) arrays.
As $H$ takes $m\Ceil{\log{\sigma}}$ bits for the quotients,  \CBT{} can represent $\Trie_\DictB$ in $m\log\sigma + \Order(m)$ expected bits of space.

\paragraph{Practical Representation of the Displacement Array}

The representation of $D$ with the CDRW array seems impractical. Poyias et al. \cite{poyias2018mbonsai} gave an alternative practical representation, where $D$ is represented by three data structures $D_1$, $D_2$ and $D_3$ as follows.

\begin{enumerate}
    \item $D_1$ is a simple array of length $m$ in which each element uses $\Delta_1$ bits for a constant $\Delta_1 > 1$.

    \item $D_2$ is a \emph{compact hash table (CHT)} described by Cleary \cite{cleary1984compact}, which stores keys from $\mathcal{U} = \{0,1,\ldots,m-1\}$ and values from $\{0,1,\ldots,2^{\Delta_2}-1\}$ for a constant $\Delta_2 > 1$.
    The keys are stored in a closed hash table of length $m' < m$ through the compact hashing technique \cite{books:knuth1998art}, where $m'$ is a power of two (a property that is in common with $m$). 
    In detail, the hash table consists of 
    \begin{itemize}
    \item a bijective transform $h : \mathcal{U} \rightarrow \mathcal{U}$, 
    \item an integer array~$Q$ of length~$m'$ to store the quotients of the keys (i.e., entry indices of $D$) representable in $\log(m/m')$ bits,
    \item an integer array~$F$ of length~$m'$ to store displacement values of $D$ representable in $\Delta_2$ bits, and
    \item two bit arrays each of length~$m'$ storing the displacement values of the quotients in $Q$ (not to be confused with the displacement values stored in $F$).
    \end{itemize}
    On inserting a key $k \in \mathcal{U}$, we store its quotient $\Floor{h(k)/m'}$ in the first vacant slot in $Q$ starting at the initial address $h(k) \bmod m'$.
    The collisions in $Q$ are therefore resolved with linear probing.
    However, this collision resolution poses the same problem as in \CBT{}, as additional displacement information is required to restore the initial address of a stored quotient in $Q$.
    Cleary solves this problem by using two bit arrays (see \cite{cleary1984compact}).
    Finally, $F[i]$ stores the value associated with the key whose quotient is stored in $Q[i]$.
    Since $F$ uses $m'\Delta_2$ bits of space,
    $D_2$ uses $m'\log (m/m') + m'\Delta_2 + 2m'$ bits of space in total.
    
    \item $D_3$ is a standard associative array that maps keys from $\mathcal{U}$ to values from $\mathcal{U}$. In our implementation, $D_3$ is a closed hash table with linear probing. Given $m''$ is the capacity of $D_3$, $D_3$ takes $2m'' \log m$ bits.
\end{enumerate}
The representation of the entry $D[i]$ for an integer~$i$ depends on its actual value:
\begin{enumerate}
\item If $D[i] < 2^{\Delta_1} - 1$, then we store $D[i]$ in the  $\Delta_1$ bits of $D_1[i]$.
\item If $2^{\Delta_1} - 1 \leq D[i] < 2^{\Delta_1} + 2^{\Delta_2}$, we represent $D[i]$ by the key-value pair $\Tuple{i, D[i] - 2^{\Delta_1} }$ stored in $D_2$.
\item Finally, if $D[i] \geq 2^{\Delta_1} + 2^{\Delta_2}$, we represent $D[i]$ by the key-value pair $\Tuple{i, D[i]}$ stored in $D_3$.
\end{enumerate}

In the experiments, we set $\Delta_1 = 4$ and $\Delta_2 = 7$.
We set the initial capacities of $D_2$ and $D_3$ to $m'=2^{12}$ and $m''=2^6$, respectively.
We set the maximum load factor of $D_2$ and $D_3$ to 0.9.
If the actual load factor of $D_2$ (resp.\ $D_3$) reaches the maximum load factor 0.9, we double the size of $D_2$ (resp.\ of $D_3$).

\paragraph{Design of the Bijective Transform}

Since we assume that $m$, $m'$, and $\sigma$ are powers of two, the bijective transform is $h: \{0,1, \ldots, 2^z - 1 \} \rightarrow \{0,1, \ldots, 2^z - 1 \}$ for some $z$.
We design this function as the concatenation of two bijective functions $h = h_1 \circ h_2$, where $h_1(x) = x \oplus \Floor{x/2^{a}}$ for an integer $a$ larger than $\Floor{z/2}$ and $h_2(x) = x p \bmod 2^z$ for a large prime $p$ smaller than $2^z$.
$h_1$ is based on the XorShift random number generators \cite{marsaglia2003xorshift}, where the inverse function $h^{-1}_1$ is given by $h^{-1}_1(x) = h_1(x)$.
The inverse function $h^{-1}_2$ of $h_2$ is given by $h^{-1}_2(x) = xp^{-1} \bmod 2^z$, where $p^{-1} \in \{1,2, \ldots, 2^z-1 \}$ is the multiplicative inverse of $p$ such that $pp^{-1} \bmod 2^z = 1$ (see \cite{koeppl2020fast} for details).
By construction, the inverse function $h^{-1}$ of $h$ is $h^{-1} = h^{-1}_2 \circ h^{-1}_1$.
Our hash function is inspired by the SplitMix algorithm \cite{steele2014fast}.

\subsection{Linear-Time Growing Algorithm}
\label{sect:bonsai:grow}

If the load factor $\alpha$ of hash table $H$ of length $m$ reaches the maximum load factor $\hat{\alpha}$, we create a new hash table $H'$ (and a new displacement array $D'$ for \CBT{}) of length $2m$ and relocate all nodes to $H'$.
Since a node depends on the position of its parent in~$H$, we can relocate a node only after having relocated all its ancestors.
This can be done in a top-down traversal (e.g., in BFS or DFS order) of the tree during which all children of a node are successively selected. 
However, because selecting all children of a node is performed by checking $\GetChild$ for \emph{all} possible characters in $\AlphB$, the relocation based on a top-down traversal needs $\Order(n\sigma)$ expected time and is therefore only for tiny alphabets practical.
Here we describe a bottom-up approach that is based on the approach by Arroyuelo et al.~\cite{arroyuelo2017lz78}.
This approach, called \emph{growing algorithm}, runs in $\Order(n)$ expected time.
A pseudo code of it is shown in \gref{alg:grow}.

Given a trie $\Trie_\DictB$ with a hash table $H$ of length $m$, the algorithm constructs an equivalent trie $\Trie'_\DictB$ with a hash table $H'$ of length $2m$.
To explain the algorithm, we define two operations $\GetEdge(u)$ returning the branching character of node $u$ and $\GetParent(u)$ returning the parent id of node $u$.
They can be computed in constant time because $H[u]$ explicitly stores the branching character and the parent id as the hash key in \PBT{}.
\CBT{} can also restore the hash key from $H[u]$ and $D[u]$.

\def\done{\textsf{Done}}
\def\map{\textsf{Map}}
\def\path{\pi}

\begin{algorithm}[tb]
\small
\caption{Linear-time growing algorithm of \PBT{} and \CBT{}}
\label{alg:grow}
\begin{algorithmic}[1]
\Require Trie $\Trie_\DictB$ with hash table $H$ of size $m$
\Ensure Equivalent trie $\Trie'_\DictB$ with hash table $H'$ of size $2m$
\State Create an empty trie $\Trie'_\DictB$ with hash table $H'$ of size $2m$ and create its root
\State Create an integer array $\map$ and a bit array $\done$, each of length $m$
\State Initialize $\done[i] \gets \Str{0}$ for all $i$
\State $\done[u_1] \gets \Str{1}$ and $\map[u_1] \gets u'_1$, where $u_1$ and $u'_1$ are the root ids of $\Trie_\DictB$ and $\Trie'_\DictB$, respectively
\For{$i = 0,\ldots,m-1$}
    \IfThen{$H[i]$ is empty}{\textbf{continue}}
    \State $u \gets i$ and $\path \gets$ empty string
    \While{$\done[u] \neq \Str{1}$} \Comment{Climb up $\Trie_\DictB$}
        \State $\path \gets \Trie_\DictB.\GetEdge(u) + \path$ \Comment{Prepend ancestor to $\path$}
        \State $u \gets \Trie_\DictB.\GetParent(u)$
    \EndWhile
    \State $u' \gets \map[u]$
    \For{$c \in \path$} \Comment{Walk down the computed path}
        \State $u \gets \Trie_\DictB.\GetChild(u,c)$ and $u' \gets \Trie'_\DictB.\AddChild(u',c)$
        \State $\map[u] \gets u'$ and $\done[u] \gets \Str{1}$
    \EndFor
\EndFor
\State \textbf{output} $\Trie'_\DictB$
\end{algorithmic}
\end{algorithm}

In the growing algorithm, we initially define two auxiliary arrays $\map$ and $\done$: 
$\map$ is an integer array and $\done$ is a bit array, each of length $m$.
We store in $\done[u]$ a \Str{1} after relocating the node stored in $H[u]$.
We keep the invariant that whenever $\done[u] = \Str{1}$, then $\map[u]$ stores the position in $H'$ of the node stored in $H[u]$.
All bits in $\done$ are initialized by $\Str{0}$ except for the root.
We scan $H$ from left to right and perform the following steps for each non-vacant slot $i$.
We first set $u$ to $i$ and $\path$ to an empty string, and then climb up the path from the node $u$ to the root.
We prematurely stop when encountering a node $v$ with $\done[v] = \Str{1}$. In this case, all ancestors of $v$ have already been relocated such that there is no need to visit them again.
Subsequently, we walk down the computed path~$\path$ while relocating the visited nodes.
Since we do not reprocess already visited nodes, we can perform the node relocation in $\Order(m) + \Order(n) = \Order(n)$ expected time, with $n = \hat{\alpha} \cdot m$ for a constant loaf factor $\hat{\alpha}$.

\paragraph{Extra Working Space}

\gref{alg:grow} maintains the auxiliary arrays $\map$ of $m\Ceil{\log{(2m)}}$ bits, $\done$ of $m$ bits and $\path$ of $h \Ceil{\log \sigma}$ bits, where $h$ is the height of $\Trie_\DictB$.
Thus, the extra working space is $m\Ceil{\log{m}} + 2m + h \Ceil{\log \sigma}$ bits if we create the auxiliary arrays naively.
However, the working space of $\map$ can be shared with $H$ because $H[i]$ for $\done[i] = \Str{1}$ is no longer needed.
In \PBT{}, the working space of $\map$ can be fully placed in $H$ because the space of $H$ is $m\Ceil{\log(m\sigma)}$ bits and $\sigma$ is at least $2$ in practice.\footnote{Even for $\sigma = 1$, a simple bit array suffices.}
Based on this in-place approach, the extra working space of \gref{alg:grow} is only $m + h \Ceil{\log \sigma}$ bits, taking account for $\done$ and $\path$ in \PBT{}.
In practice, the space of $\path$ is negligible because $h$ is bounded by the maximum length of keywords in $\Dict$ and $h \ll m$.

In \CBT{}, $H$ uses only $m\Ceil{\log\sigma}$ bits.
As $\sigma \ll m$ in most scenarios, it is difficult to completely store $\map$ in $H$; however, we can also use the space of $D_1$, which is $m\Delta_1$ bits.
If $\Ceil{\log{(2m)}} \leq \Ceil{\log\sigma} + \Delta_1$, $\map$ can be fully placed in $H$ and $D$; otherwise, the extra working space of $m(\Ceil{\log{(2m)}} - \Ceil{\log\sigma} - \Delta_1)$ bits for $\map$ is needed in addition to that of $\done$ and $\path$.

\subsection{NLM Data Structures}
\label{sect:bonsai:nlm}

In m-Bonsai, the node ids are values drawn from the universe $[0,m)$ whose randomness depend on the used hash function.
As the task of an NLM data structure is to map node ids to their respective node labels, an appropriate NLM data structure for m-Bonsai is a dynamic associative array that stores node label strings $L_i$ for arbitrary integer keys $i \in [0,m)$.
In what follows, we first present a plain approach and then show how to compactify it.

\paragraph{Plain NLM}

\begin{figure*}[tb]
\centering
\subfloat[\PLM{}]{
    \includegraphics[scale=\FigScale]{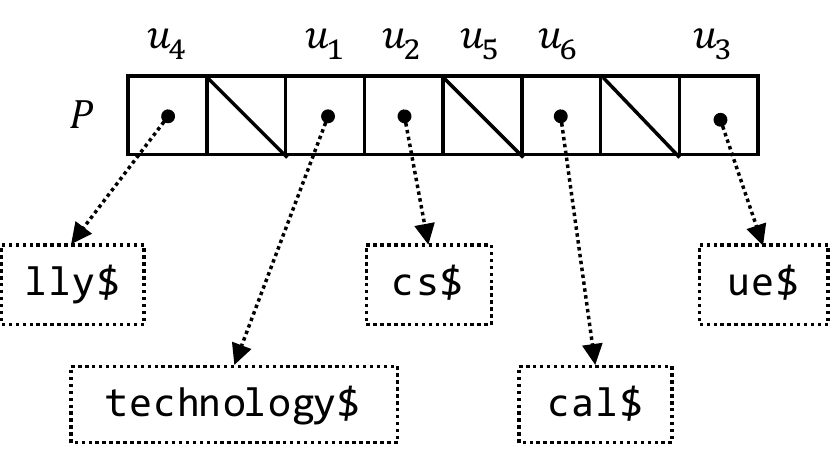}
    \label{fig:bt-plm}
}
\subfloat[\CLM{}]{
    \includegraphics[scale=\FigScale]{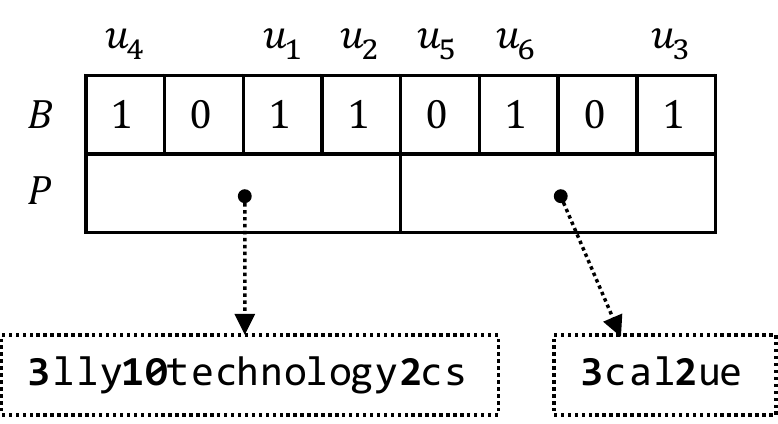}
    \label{fig:bt-clm}
}
\caption{Examples of NLM in m-Bonsai for the DynPDT in \fref{fig:ipd-e}.}
\label{fig:bt-nlm}
\end{figure*}

The simplest approach is to use a pointer array $P$ of length $m$ such that $P[i]$ stores the pointer to $L_i$ or $\False$ if no node with id $i$ exists.
We refer to the approach as \PLM{} (Plain Label Map).
\fref{fig:bt-plm} shows an example of \PLM{}.
Given a node of id $i$, \PLM{} can obtain $L_i$ through $P[i]$ in $\Order(1)$ time.
However, $P$ takes $mw = \Order(m \log m)$ bits, where the word size is $w = \Theta(\log m)$.
This space consumption is obviously large.

\paragraph{Sparse NLM}

We present an alternative compact approach that reduces the pointer overhead of \PLM{} in a manner similar to Google's sparse hash table \cite{software:sparsehash}.
In this approach, we divide the node labels into \textit{groups} of $\ell = \Theta (w)$ labels over the ids.
That is, the first group consists of $L_0 , L_1 \ldots , L_{\ell - 1}$, the second group consists of $L_{\ell}, L_{\ell+1}, \ldots , L_{2\ell - 1}$, and so on.
Moreover, we introduce a bitmap $B$ such that $B[i] = \Str{1}$ iff $L_i$ exists.
We concatenate all node labels $L_i$ with $B[i] = \Str{1}$ of the same group together, sorted in the id order.
The length of $P$ becomes $\Ceil{m/\ell}$ by maintaining, for each group, a pointer to its concatenated label string.
We refer to the approach as \CLM{} (Sparse Label Map).

With the array $P$ and the bitmap $B$, we can access $L_i$ as follows:
If $B[i] = \Str{0}$, we are done since $L_i$ does not exist in this case; otherwise, we obtain the concatenated label string storing $L_i$ from $P[g]$, where $g = \Floor{i/\ell}$.
Given $j = \sum_{k=0}^{i \bmod \ell}B_g[k]$ for the bit chunk $B_g := B[g \ell, (g+1) \ell)$, $L_i$ is the $j$-th node label of the concatenated label string.
As $\ell = \Theta (w)$, counting the occurrences of \Str{1}s in chunk $B_g$ is supported in constant time using the \emph{popcount} operation \cite{gonzalez2005practical}.
It is left to explain how to search $L_i$ in the respective concatenated label string.
For that we present two representations of the concatenated label strings:

\begin{enumerate}
\item If the node labels are straightforwardly concatenated (e.g., the second group in \fref{fig:bt-plm} is \Str{cal\$ue\$} in $\ell = 4$), we can sequentially count the $\texttt{\$}$ delimiters to find the $(j-1)$-th delimiter marking the ending of the $(j-1)$-th stored string, after which $L_i$ starts.
We can therefore extract $L_i$ in $\Order(\ell \Lambda)$ time, where $\Lambda$ again denotes the maximum length of all node labels.
\item We can shorten the scan time with the \emph{skipping} technique used in array hashing \cite{askitis2005cache}.
This technique puts its length in front of each node label via some prefix encoding such as VByte \cite{williams1999compressing}.
Note that we can omit the terminators of each node label.
The skipping technique allows us to jump ahead to the start of the next node label; therefore, the scan is supported in $\Order(\ell)$ time.
\fref{fig:bt-clm} shows an example of \CLM{} with the skipping technique.
\end{enumerate}

Regarding the space usage of \CLM{}, $P$ and $B$ use $w\Ceil{m/\ell}$ and $m$ bits, respectively.
For $\ell = \Theta (w)$, the total space usage becomes $\Order(m)$ bits, which is smaller than $mw$ bits in \PLM{}; however, the access time is $\Order(w) = \Order(\log m)$.

\section{Representation Based on FK-hash}
\label{sect:fkhash}

This section presents our DynPDT representation approaches based on FK-hash \cite{fischer2017practical}.
The basic idea of FK-hash is the same as that of m-Bonsai.
The difference is that FK-hash incrementally assigns node ids and explicitly stores them as values in the hash table, while m-Bonsai uses the locations of the stored elements of the hash table as node ids.
Although FK-hash uses more space than m-Bonsai, the assignment of node ids simplifies the growing algorithm. 

\paragraph{Outline}
In the same manner as m-Bonsai, we consider a plain and a compact representation based on FK-hash.
In \sref{sect:fkhash:rep} we present both representations.
In \sref{sect:fkhash:nlm} we propose NLM data structures designed for FK-hash.

\subsection{Trie Representations}
\label{sect:fkhash:rep}

Like m-Bonsai, FK-hash locates nodes on a closed hash table $H$ of length $m$, but does not use the addresses of $H$ as node ids.
FK-hash incrementally assigns node ids from zero and explicitly stores them in an integer array $M$ of length $m$.
In other words, when creating the $u$-th node by storing it in $H[i]$, its node id is $u$, which is stored in $M[i]$.
In a way similar to m-Bonsai, $\AddChild(u,c)$ is performed as follows:
We compose the key $k = \Tuple{u,c}$, hash it with~$h$, and then search the first vacant slot $H[i']$ from $i = h(k) \bmod m$ by linear probing.
Given $u_{\text{max}}$ is the currently largest node id, we assign the id $v = u_{\text{max}} + 1$ to the new child, and set $H[i'] = k$ and $M[i'] = v$.
The displacement information $i' - i$ is maintained analogously to m-Bonsai.

In the same manner as m-Bonsai, we can think of two representations depending on whether $H$ is compactified or not.
The non-compact one is referred to as \PHT{} (Plain FK-hash Trie).
The compact one is referred to as \CHT{} (Compact FK-hash Trie).
Compared to \PBT{} and \CBT{}, \PHT{} and \CHT{} keep an additional integer array $M$ and require $m \Ceil{\log{n}}$ additional bits of space.

\paragraph{Table Growing}

An advantage of FK-hash is that growing the hash table is done in the same manner as in standard closed hash tables.
In detail, $H$ can be enlarged by scanning nodes on $H$ from left to right and relocating the nodes in a new hash table $H'$ of length $2m$.
The growing algorithm takes $\Order(m)$ expected time.
This time complexity is identical to that of \gref{alg:grow}; however, the growing algorithm of FK-hash is faster in practice because of its simplicity.
In addition, no auxiliary data structure is needed like $\map$ and $\done$ used by \gref{alg:grow}.

\subsection{NLM Data Structures}
\label{sect:fkhash:nlm}

\begin{figure*}[tb]
 \centering
 \subfloat[\PLM{}]{
    \includegraphics[scale=\FigScale]{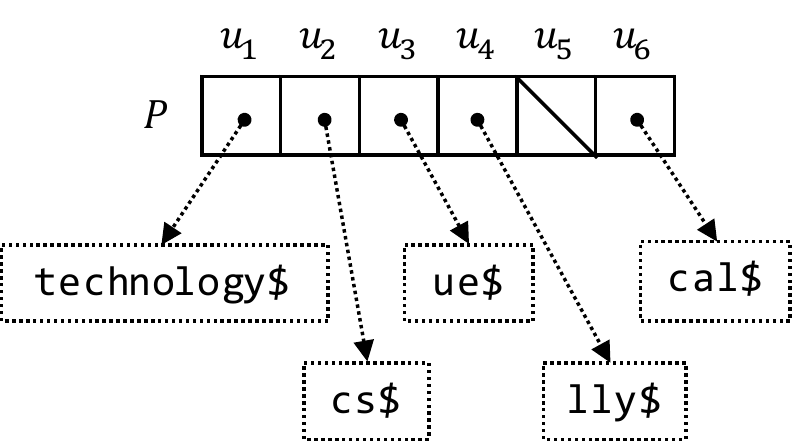}
    \label{fig:fkt-plm}
 }
 \subfloat[\CLM{}]{
    \includegraphics[scale=\FigScale]{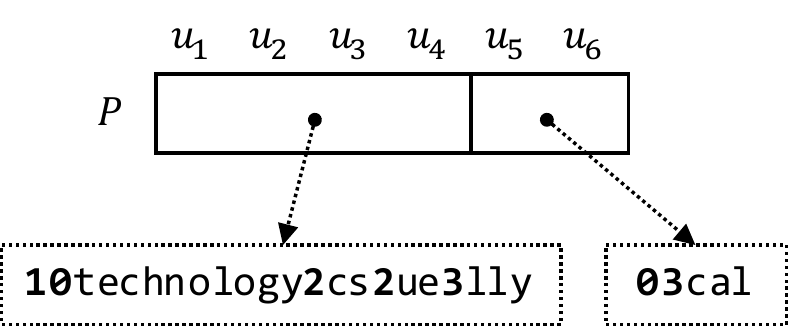}
    \label{fig:fkt-clm}
 }
\caption{Examples of NLM in FK-hash for the DynPDT in \fref{fig:ipd-e}.}
 \label{fig:fkt-nlm}
\end{figure*}

Like in \sref{sect:bonsai:nlm}, we introduce \PLM{} and \CLM{} adapted to FK-hash.
\fref{fig:fkt-nlm} shows an example for each of them.
Although \PLM{} in FK-hash is basically identical to that in m-Bonsai, \CLM{} can be simplified as follows.

In m-Bonsai, it is necessary to identify whether $L_i$ exists and the rank of $L_i$ in the group because node ids are randomly assigned; therefore, we introduced a bitmap $B$ of length $m$ and utilized the popcount operation.
In FK-hash, however, such a bitmap is not needed because node ids are incrementally assigned.
Put simply, a node label $L_i$ is stored in the group of id $g = \Floor{i/\ell}$ and located at the $(i \bmod \ell)$-th position in the group.
When using the skipping technique,
care has to be taken for the step nodes whose node labels are empty. For each of them, we put the length 0 in its corresponding concatenated label string.
For example, we put a '0' in the second concatenated label string for the step node $u_5$ in \fref{fig:fkt-clm}.
Finally, we can insert a new node label by appending it to the last concatenated label string.

\section{Experiments}
\label{sect:exp}

In this section we evaluate the practical performance of DynPDT.
The source code for our experiments are available at \url{https://github.com/kampersanda/dictionary_bench}.

\subsection{Setup}

We conducted all experiments on one core of a quad-core Intel Xeon CPU E5-2680 v2 clocked at 2.80 Ghz in a machine with 256 GB of RAM, running the 64-bit version of CentOS 6.10 based on Linux 2.6.
We implemented our data structures in C++17.
We compiled the source code with g++ (version 7.3.0) in optimization mode -O3.
We used 4-byte integers for the values associated with the keywords.

\paragraph{Datasets}

Our benchmarks are based on the following eight real-world datasets:

\begin{itemize}
    \item \Geo{} consists of 7 million different names for the geographic points provided by the GeoNames database.\footnote{\url{http://download.geonames.org/export/dump/}} Managing such geographic identifiers within a limited resource is essential in modern geographic information systems as described in \cite{martinez2016practical}. We obtained the geographic names by extracting the \emph{asciiname} column of the GeoNames dump in the same manner as \cite{martinez2016practical}.
    \item \AOL{} consists of 10 million different search queries in the AOL database, which is a huge collection of 20 million search queries from 650,000 users sampled over three months.\footnote{\url{http://www.cim.mcgill.ca/~dudek/206/Logs/AOL-user-ct-collection/}} The dataset contains keywords written in natural English, which has been often used to benchmark search algorithms such as \cite{grossi2014fast}.
    \item \Wiki{} consists of 14 million different page titles from the English Wikipedia dump at September 2018.\footnote{\url{https://dumps.wikimedia.org/enwiki/}} As the dataset contains various special characters encoded in UTF-8, the alphabet size is larger than that of AOL. It is also a well-used dataset to benchmark search algorithms such as \cite{grossi2014fast,arz2018lempel,kanda2017compressed}.
    \item \DNA{} consists of all 12-mers (i.e., substrings of length 12) found in the DNA dataset from the Pizza\&Chili corpus.\footnote{\url{http://pizzachili.dcc.uchile.cl/texts/dna/}} Among the used datasets, it has the smallest alphabet and the shortest keywords. The number of keywords is 15 million. In bioinformatics, popular alignment software need to manage such keywords within limited space as described in \cite{martinez2016practical}.
    \item \LUBMS{} consists of 53 million different URIs extracted from the RDF dataset generated by the Lehigh University Benchmark \cite{guo2005lubm} for 1,600 universities.\footnote{The dataset is distributed under the name `DS5' at \url{https://exascale.info/projects/web-of-data-uri/}.} Modern RDF systems \cite{wylot2011diplodocus,wylot2014tripleprov} encode URIs in a huge set into unique integers by using a dynamic keyword dictionary. The dataset is evaluated in \cite{mavlyutov2015comparison} to analyze the performances of RDF systems.
    \item \LUBML{} consists of 230 million different URIs extracted from the RDF dataset generated by the Lehigh University Benchmark \cite{guo2005lubm} for 7,000 universities.\footnote{Although this dataset is not distributed, one can obtain the identical dataset through the LUBM data generator (called UBA) at \url{http://swat.cse.lehigh.edu/projects/lubm/}.} The dataset is a larger version of \LUBMS{}. It is also evaluated in \cite{mavlyutov2015comparison}.
    \item \UK{} consists of 40 million different URLs obtained from a 2005 crawl of the \texttt{.uk} domain performed by UbiCrawler \cite{boldi2004ubicrawler}.\footnote{\url{http://law.di.unimi.it/webdata/uk-2005/}} URLs are traditionally used to benchmark search algorithms for long strings such as \cite{grossi2014fast,arz2018lempel,kanda2017compressed,askitis2010engineering}. Also, the modern Web crawler \cite{ueda2013parallel} manages a huge set of URLs by using a dynamic keyword dictionary.
    \item WebBase consists of 118 million different URLs of a 2001 crawl performed by the WebBase crawler \cite{hirai2000webbase}.\footnote{\url{http://law.di.unimi.it/webdata/webbase-2001/}} The dataset is larger than \UK{} and also used in previous experiments of keyword dictionaries such as \cite{grossi2014fast}.

\end{itemize}

\begin{table}
\small
\centering
\caption{Statistics for the datasets used. \Size{} is the total length of the keywords, $n$ is the number of all distinct keywords in millions (M), \MinLen{} (resp. \MaxLen{} and \AveLen{}) is the maximum (resp. minimum and average) length of the keywords, and $|\AlphA|$ is the actual alphabet size of the keywords.}
\label{tab:dataset}
\begin{tabular}{lrrrrrr}
\toprule
 & \Size{} & $n$ & \MinLen{} & \MaxLen{} & \AveLen{} & $|\AlphA|$ \\
\midrule
\Geo{} & 109 MiB & 7.3 M & 2 & 152 & 15.7 & 99 \\
\AOL{} & 224 MiB & 10.2 M & 2 & 523 & 23.2 & 85 \\
\Wiki{} & 286 MiB & 14.1 M & 2 & 252 & 21.2 & 200 \\
\DNA{} & 189 MiB & 15.3 M & 13 & 13 & 13.0 & 16 \\
\LUBMS{} & 3.1 GiB & 52.6 M & 10 & 80 & 63.7 & 57 \\
\LUBML{} & 13.8 GiB & 230.1 M & 10 & 80 & 64.2 & 57 \\
\UK{} & 2.7 GiB & 39.5 M & 17 & 2,030 & 72.4 & 103 \\
\WebBase{} & 6.6 GiB & 118.2 M & 10 & 10212 & 60.2 & 223 \\
\bottomrule
\end{tabular}

\end{table}

\tref{tab:dataset} summarizes relevant statistics for each dataset.

\subsection{Average Height}
\label{sect:exp:height}

We evaluate the average height of the DynPDT $\PDT_\Dict$ built on our datasets.
The average height of $\PDT_\Dict$ is the arithmetic mean of the heights of all nodes over the number of nodes, omitting step nodes in the calculation.
Although the average height is an important measure related to the average number of random accesses, we cannot \textit{a priori} predict the average height of DynPDT because this number depends on the insertion order of the keywords.
To reason about the quality of the average height, we study it in relation to the following known lower and upper bounds on it:
The lower bound is the average height of the path-decomposed trie created by  the centroid path decomposition \cite[Corollary 3]{daigle2016optimal}.
The upper bound is the average height of the path-decomposed trie created by always choosing the child whose subtrie has the \emph{fewest} number of leaves.

\begin{table}[tb]
\small
\centering
\caption{Experimental results of the average heights of $\PDT_\Dict$ and $\Trie_\Dict$ denoted by \AveHeight{}. Also, \AveHeightLB{} and \AveHeightUB{} are the lower bound and the upper bound of the average height of $\PDT_\Dict$, respectively (defined in \sref{sect:exp:height}). \AveHeightLB{} is the average height of the path-decomposed trie obtained by the centroid path decomposition. \AveHeightUB{} is the average height of the path-decomposed trie obtained by the path decomposition selecting children with the fewest leaves.}
\label{tab:height}
\begin{tabular}{lrrrrrrrr}
\toprule
 & \Geo{} & \AOL{} & \Wiki{} & \DNA{} & \LUBMS{} & \LUBML{} & \UK{} & \WebBase{} \\
\midrule
\AveHeight{} of $\PDT_\Dict$ & 6.0 & 6.2 & 6.3 & 9.0 & 7.5 & 7.9 & 7.8 & 7.3 \\
\AveHeightLB{} of $\PDT_\Dict$ & 5.2 & 5.2 & 5.3 & 8.9 & 6.6 & 7.4 & 6.0 & 6.2 \\
\AveHeightUB{} of $\PDT_\Dict$ & 8.5 & 10.5 & 9.7 & 10.7 & 11.8 & 12.4 & 14.7 & 15.4 \\
\midrule
\AveHeight{} of $\Trie_\Dict$ & 15.7 & 23.2 & 21.2 & 13.0 & 63.7 & 64.2 & 72.4 & 60.2 \\
\bottomrule
\end{tabular}

\end{table}

\tref{tab:height} shows the experimental results of the average heights of $\PDT_\Dict$ and $\Trie_\Dict$ for all the datasets.
To analyze the performance of DynPDT in our experiments, we constructed DynPDT dictionaries by inserting keywords in random order.
For that, we shuffled the dataset with the Fisher--Yates shuffle algorithm \cite{durstenfeld1964algorithm}.
Naturally, the actual average heights of $\PDT_\Dict$ are between their lower and upper bounds, and those of $\Trie_\Dict$ are the same as \AveLen{}.
The upper bounds are more than twice as large as the lower bounds for \AOL{}, \UK{}, and \WebBase{}; however, the upper bounds were up to 5.4x smaller than the average heights of $\Trie_\Dict$ due to the path decomposition, especially for long keywords such as URIs.
Therefore, the incremental path decomposition can make dynamic keyword dictionaries more cache-friendly, especially for long keywords even if the insertion order is inconvenient and the average height is close to the upper bound.

\subsection{Parameter for Step Nodes}
\label{sect:exp:step}

The parameter $\lambda$ influences the number of step nodes.
We analyze the space and time performance of DynPDT when varying the parameter $\lambda$.
In this experiment, we constructed DynPDT dictionaries for each parameter $\lambda \in \{4,8,16,\ldots,1024\}$ on the datasets \Wiki{}, \LUBMS{} and \UK{}, and observed the working space and the construction time.
For the DynPDT representation, we tested the combination of \CHT{} and \CLM{}  with $\ell = 16$, referred to as \CCHT{} in the following.
As described in \sref{sect:exp:height}, the dictionary was constructed by inserting keywords in random order.
The working space was measured by checking the maximum resident set size (RSS) required during the online construction.

\begin{table}[tb]
\small
\centering
\caption{Experimental results of \CCHT{} for various values of the parameter $\lambda$. \Steps{} is the proportion of the number of step nodes among all nodes in DynPDT, \Space{} is the working space in GiB, and \Time{} is the elapsed time for the construction in seconds.}
\label{tab:lambda}
\begin{tabular}{rrrrrrrrrr}
\toprule
 & \multicolumn{3}{c}{\Wiki{}} & \multicolumn{3}{c}{\LUBMS{}} & \multicolumn{3}{c}{\UK{}} \\
\cmidrule(lr){2-4}
\cmidrule(lr){5-7}
\cmidrule(lr){8-10}
$\lambda$ & \Steps{} & \Space{} & \Time{} & \Steps{} & \Space{} & \Time{} & \Steps{} & \Space{} & \Time{} \\
\midrule
4 & 19.55\% & 0.36 & 20.9 & 7.75\% & 0.78 & 116.1 & 37.60\% & 1.22 & 116.5 \\
8 & 6.12\% & 0.27 & 18.2 & 2.83\% & 0.78 & 96.8 & 15.51\% & 1.20 & 96.2 \\
16 & 1.32\% & 0.27 & 17.1 & 0.31\% & 0.78 & 84.9 & 5.22\% & 1.20 & 87.6 \\
32 & 0.12\% & 0.27 & 17.3 & 0.02\% & 0.79 & 83.3 & 1.31\% & 1.20 & 86.0 \\
64 & 0.00\% & 0.27 & 17.2 & 0.00\% & 0.80 & 82.9 & 0.23\% & 1.21 & 85.4 \\
128 & 0.00\% & 0.27 & 17.4 & 0.00\% & 0.80 & 82.9 & 0.04\% & 1.22 & 85.3 \\
256 & 0.00\% & 0.28 & 17.2 & 0.00\% & 0.81 & 83.6 & 0.01\% & 1.22 & 85.2 \\
512 & 0.00\% & 0.28 & 17.2 & 0.00\% & 0.82 & 83.3 & 0.00\% & 1.23 & 85.4 \\
1024 & 0.00\% & 0.28 & 17.3 & 0.00\% & 0.83 & 83.1 & 0.00\% & 1.24 & 85.4 \\
\bottomrule
\end{tabular}

\end{table}

\tref{tab:lambda} shows the experimental results for construction.
Since $\lambda$ has a direct impact on $\sigma$, which influences the space usage of $H$, the working space depends on the value of $\lambda$.
Although this dependency looks like $\lambda$ and the taken space are in direct correlation, for \Wiki{} and \UK{}, the working spaces for $\lambda = 4$ (i.e., 0.36 GiB and 1.22 GiB respectively) were not the smallest.
For \Wiki{}, the reason for this is that many step nodes raised the load factor $\alpha$ and involved an additional enlargement of the hash table.
Specifically, the enlargements were conducted nine times with $\lambda = 4$, although they were conducted eight times with $\lambda \geq 8$.
For \UK{}, this reason is that the high load factor $\alpha$ caused by a huge number of step nodes raised the average displacement value stored in $D$ and involved the use of $D_2$ and $D_3$, although no additional enlargement was conducted.
Regarding the time performance, this huge number of step nodes slowed down the construction.
Therefore, a too small parameter $\lambda$ can involve large space requirements and long construction times.
On the other hand, when $16 \leq \lambda$, the working space and construction time do not significantly vary.

\begin{table}[tb]
\small
\centering
\caption{Proportion of the number of step nodes to the total number of nodes in DynPDT. Bold font indicates the results with the smallest $\lambda$ such that \Steps{} is less than 1\%. \AveNLL{} is the average length of the node labels.}
\label{tab:lambda-all}
\begin{tabular}{lrrrrrrr}
\toprule
 & \multicolumn{6}{c}{\Steps{}} & \\
\cmidrule(lr){2-7}
 & $\lambda = 4$ & $\lambda = 8$ & $\lambda = 16$ & $\lambda = 32$ & $\lambda = 64$ & $\lambda = 128$ & \AveNLL{} \\
\midrule
\Geo{} & 6.34\% & 1.44\% & \textbf{0.28\%} & 0.04\% & 0.00\% & 0.00\% & 6.1 \\
\AOL{} & 22.16\% & 6.83\% & 1.26\% & \textbf{0.11\%} & 0.01\% & 0.00\% & 10.6 \\
\Wiki{} & 19.55\% & 6.12\% & 1.32\% & \textbf{0.12\%} & 0.00\% & 0.00\% & 8.7 \\
\DNA{} & \textbf{0.11\%} & 0.00\% & 0.00\% & 0.00\% & 0.00\% & 0.00\% & 1.4 \\
\LUBMS{} & 7.75\% & 2.83\% & \textbf{0.31\%} & 0.02\% & 0.00\% & 0.00\% & 3.7 \\
\LUBML{} & 7.66\% & 2.80\% & \textbf{0.31\%} & 0.02\% & 0.00\% & 0.00\% & 3.7 \\
\UK{} & 37.60\% & 15.51\% & 5.22\% & 1.31\% & \textbf{0.23\%} & 0.04\% & 18.0 \\
\WebBase{} & 24.15\% & 8.94\% & 2.46\% & \textbf{0.50\%} & 0.08\% & 0.02\% & 11.1 \\
\bottomrule
\end{tabular}

\end{table}

From this observation, we derive two facts for $\lambda$:
On the one hand, the most important recommendation is not to choose a parameter $\lambda$ that is too small.
On the other hand, choosing a large parameter $\lambda$ is not a significant problem because the space and time performance do not significantly decrease as $\lambda$ grows.
For example, when $\lambda = 32$ on \Wiki{}, the proportion of step nodes is 0.12\%; however, even with a larger parameter $\lambda$ such as 512 or 1024, the working space and construction time are almost the same.
\tref{tab:lambda-all} shows \Steps{} for each parameter $\lambda$ and the average length of the node labels (denoted by \AveNLL{}) for all the datasets.
Even for long keywords like URLs (i.e., \UK{}), \AveNLL{} is bounded by 18.0 and \Steps{} is within 1\% of all nodes when $\lambda = 64$.
Among the tested values for $\lambda$, we suggest setting $\lambda$ to 32 or 64 for keywords whose length is not much longer than that of the URL datasets.

\subsection{Comparison among DynPDT Representations}
\label{sect:exp:rep}

We compared the performance of our DynPDT representations,
for which we benchmarked the following six combinations:

\begin{itemize}
    \item \PPBT{} is the combination of \PBT{} and \PLM{},
    \item \PCBT{} is the combination of \PBT{} and \CLM{},
    \item \CCBT{} is the combination of \CBT{} and \CLM{},
    \item \PPHT{} is the combination of \PHT{} and \PLM{},
    \item \PCHT{} is the combination of \PHT{} and \CLM{}, and
    \item \CCHT{} is the combination of \CHT{} and \CLM{}.
\end{itemize}

We evaluated the working space during the construction and the running times of \Insert{} and \Lookup{}.
Like in \sref{sect:exp:step}, we constructed each dictionary and measured its working space.
To measure the \Lookup{} time, we chose 1 million random keywords from each dataset.
The running times are the average of 10 runs.
For \CLM{}, we tested $\ell \in \{8, 16, 32,64\}$.
For $\lambda$, we chose the smallest value among those from \tref{tab:lambda-all} where \Steps{} is less than 1\%.

\begin{figure}[tb]
    \centering
     \includegraphics[width=\PlotWidth]{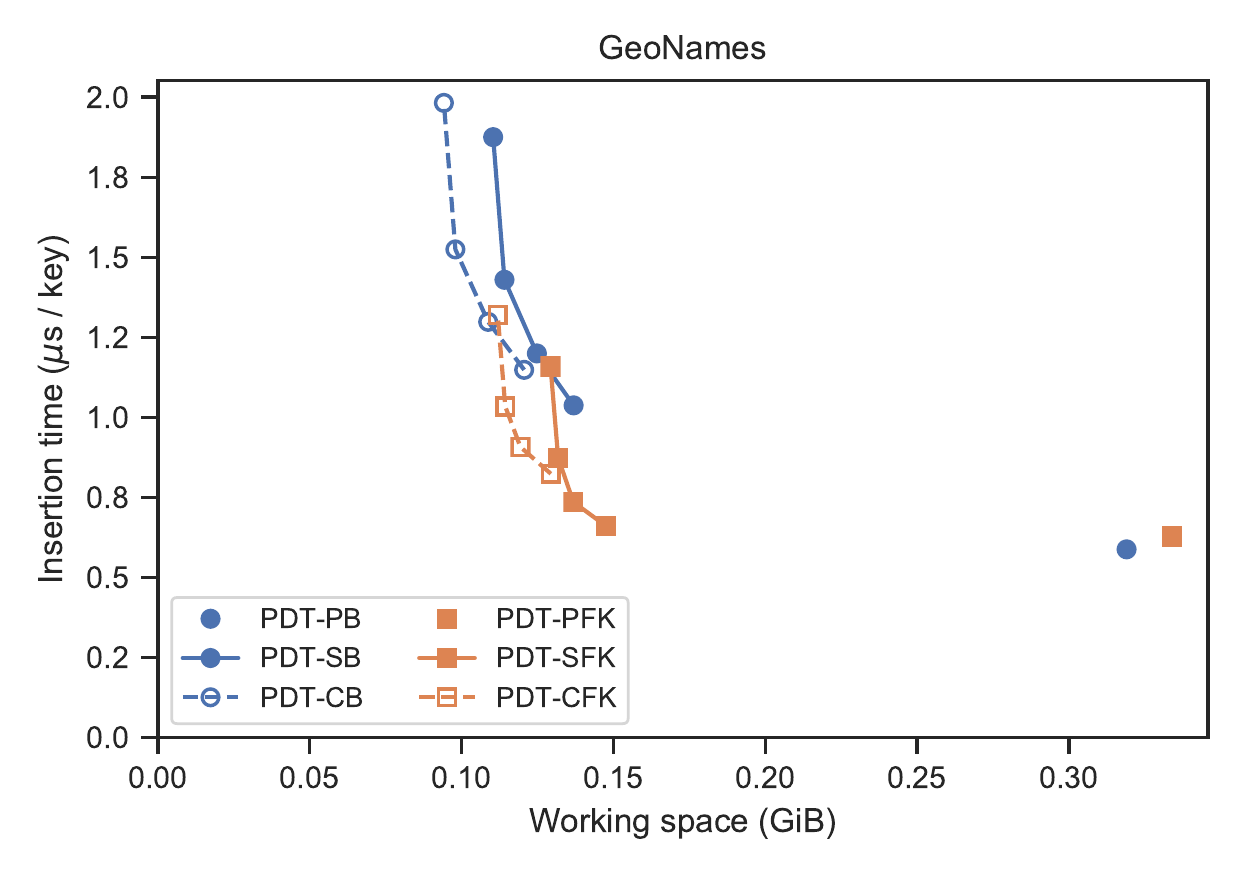}
    \includegraphics[width=\PlotWidth]{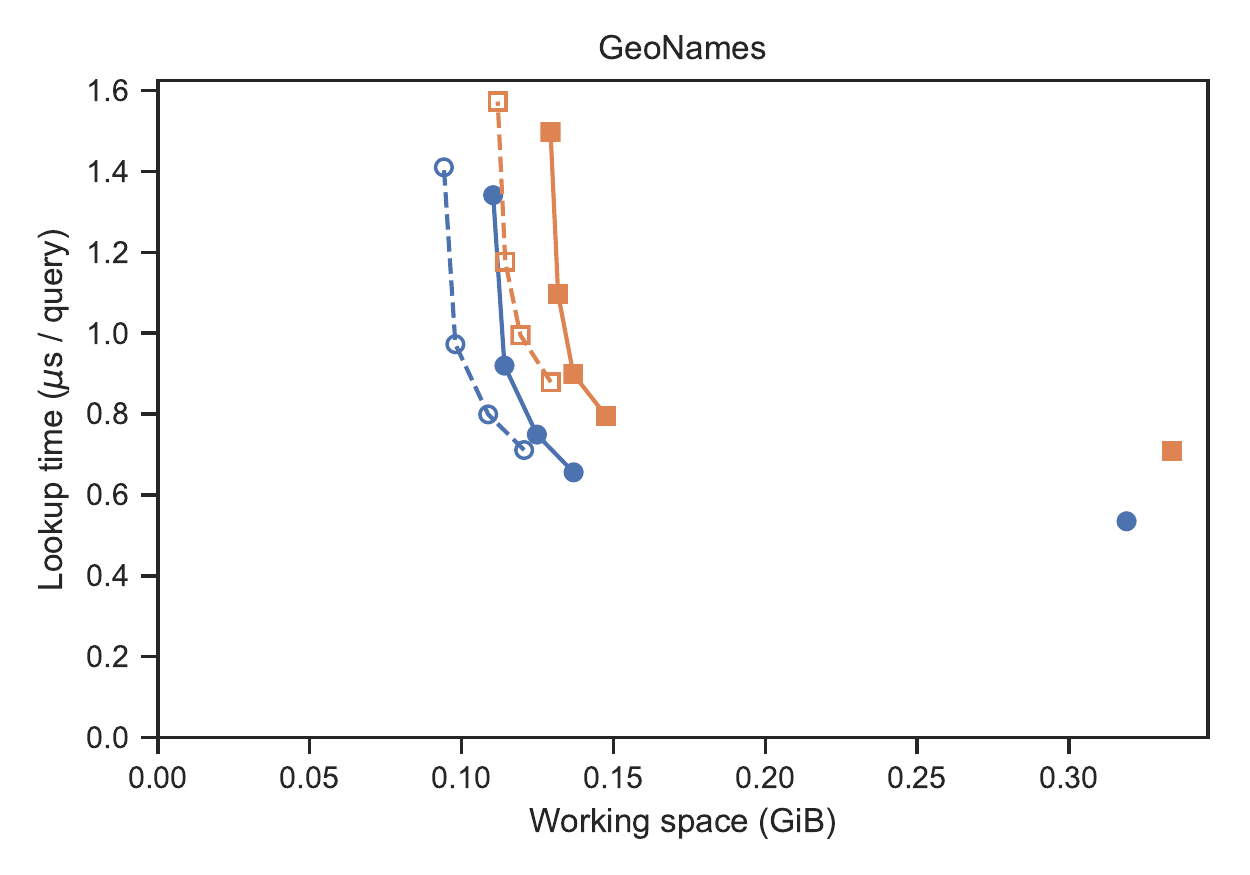}\\
    \includegraphics[width=\PlotWidth]{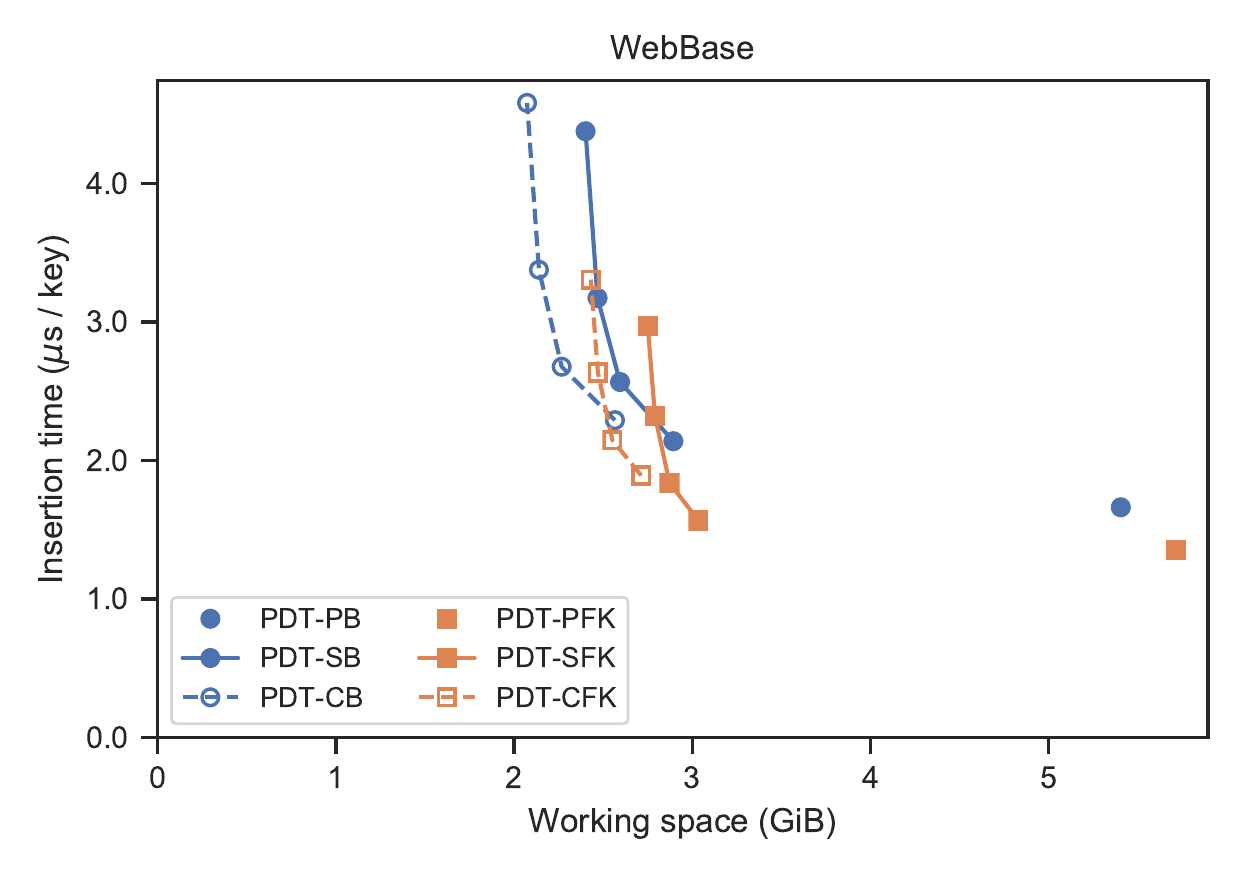}
    \includegraphics[width=\PlotWidth]{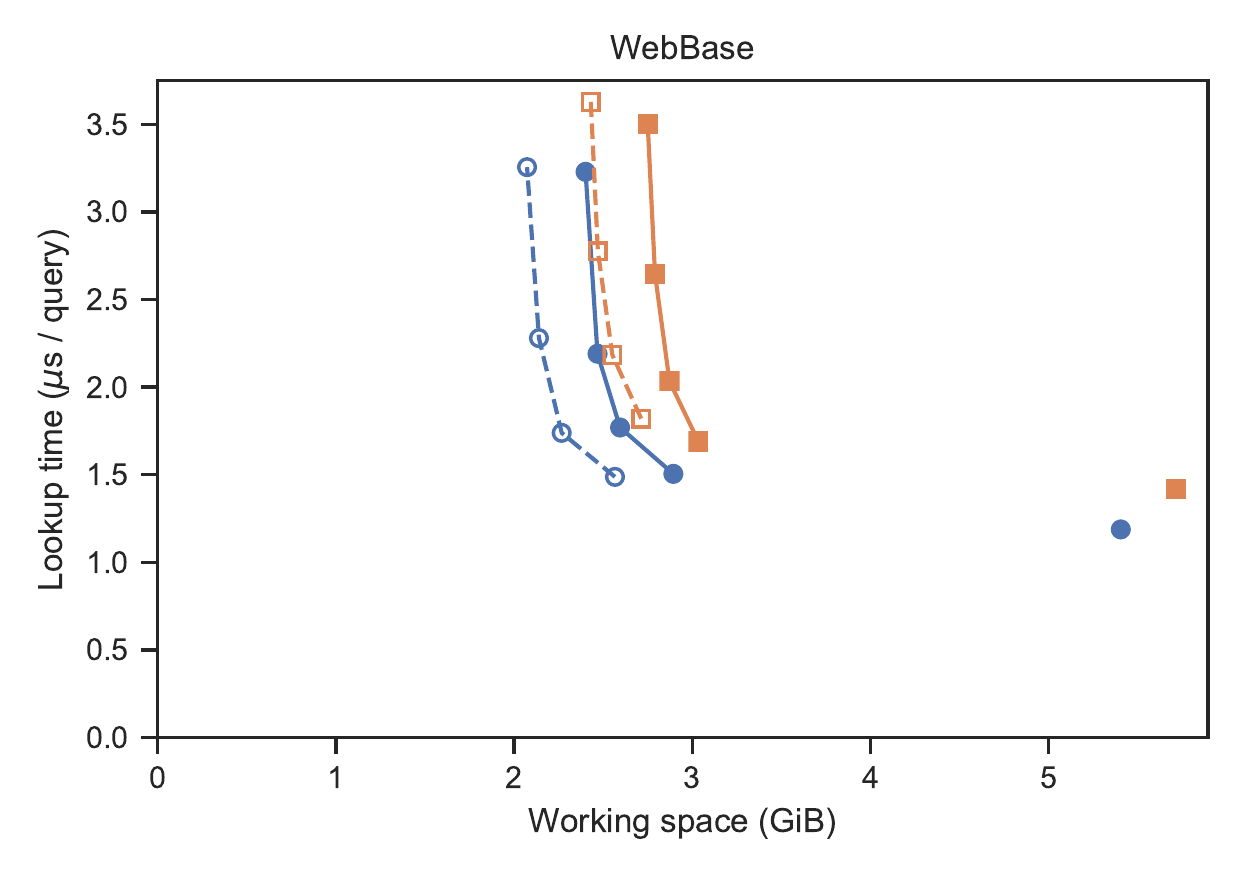}
    \caption{Experimental results for combinations of DynPDT representations.}
    \label{fig:results-comb}
\end{figure}

\fref{fig:results-comb} shows the experimental results for \Geo{} and \WebBase{}.
Regarding the representations using \CLM{}, the working space is the largest but the running times are the shortest with $\ell = 8$, and vice versa with $\lambda = 64$.
In other words, for each representation in the plots, the rightmost and lowest result is the one with $\ell = 8$, and the leftmost and highest result is the one with $\ell = 64$.

We observe that

\begin{itemize}
    \item \CLM{} significantly reduces the working space of \PLM{}. Compared to \PPBT{}, \PCBT{} is 57--65\% smaller for \Geo{} and 46--56\% smaller for \WebBase{}. Compared to \PPHT{}, \PCHT{} is 56--61\% smaller for \Geo{} and 47--52\% smaller for \WebBase{}.
    \item Regarding the representations based on m-Bonsai, the \Insert{} time of \CLM{} is slower than that of \PLM{} because inserting a new node label into the group is costly. When $\ell = 8$, the insertion of \PCBT{} is 29-163\% slower than that of \PPBT{}; however, the \Lookup{} times are competitive.
    \item Regarding the representations based on FK-hash, \CLM{} with $\ell=8$ is competitive to \PLM{} with respect to the \Insert{} time because the update algorithm is simple. Also, the \Lookup{} times are competitive.
    \item The time performance of \CLM{} with large group sizes ($\ell = 32$ or $64$) is  worse than that of \CLM{} with small group sizes ($\ell = 8$ or $16$). For example, for \Geo{}, \PCBT{} with $\ell=64$ is 19\% smaller but 81--105\% slower than \PCBT{} with $\ell=8$.
    \item The compact trie representations \CBT{} and \CHT{} are more lightweight but slower than the plain representations \PBT{} and \PHT{}; however, the differences are small. For example, \PCBT{} is 12\% smaller but 8--11\% slower than \CCBT{} for \Geo{}.
    \item The representations based on m-Bonsai are smaller than those based on FK-hash. Also regarding the \Lookup{} time, the m-Bonsai representations are faster. However, regarding the \Insert{} time, the FK-hash representations are faster because the growing algorithm is simple.
\end{itemize}

\subsection{Comparison with Existing Data Structures}
\label{sect:exp:exist}

We compare the performance of DynPDT with existing data structures.
We exhaustively tested existing implementations of dynamic keyword dictionaries such as open-source dynamic hash containers \cite{software:sparsepp,software:robin-map,software:hopscotch-map} and recent dynamic trie indexes \cite{tsuruta2020ctrie,takagi2016packed}.
However, compared to DynPDT, most of them consumed significantly more space.
For our benchmarks, we selected the following four space-efficient implementations:\footnote{All the experimental results are shown in \aref{apdx:exp}.}

\begin{itemize}
    \item \ArHash{} is a cache-conscious hash table with string keys \cite{askitis2005cache}.
    \item \HAT{} is a hybrid data structure of the burst trie \cite{heinz2002burst} and \ArHash{} \cite{askitis2010engineering}.
    \item \Judy{} is a trie-based dictionary implementation developed at Hewlett-Packard Research Labs \cite{manual:judy10min}.
    \item \Cedar{} developed by Yoshinaga \cite{yoshinaga2014self} is an efficient dictionary implementation based on dynamic double-array tries \cite{aoe1989efficient}.
\end{itemize}

For \ArHash{} and \HAT{}, we used Tessil's implementations \cite{software:array-hash,software:hat-trie}.
From the three implementation variations of \Cedar{},
we took one based on a reduced trie \cite{yoshinaga2014self} and one based on prefix trie \cite{aoe1989efficient}, and denote them by \CedarR{} and \CedarP{}, respectively.
\CedarR{} is suitable for short keywords\footnote{We cannot be more concrete here since the efficiency of the heuristics of these data structures do not merely depend on the keyword lengths.}, whereas \CedarP{} is suitable for the general case.

\begin{figure}[p]
    \centering
    \includegraphics[width=\PlotWidth]{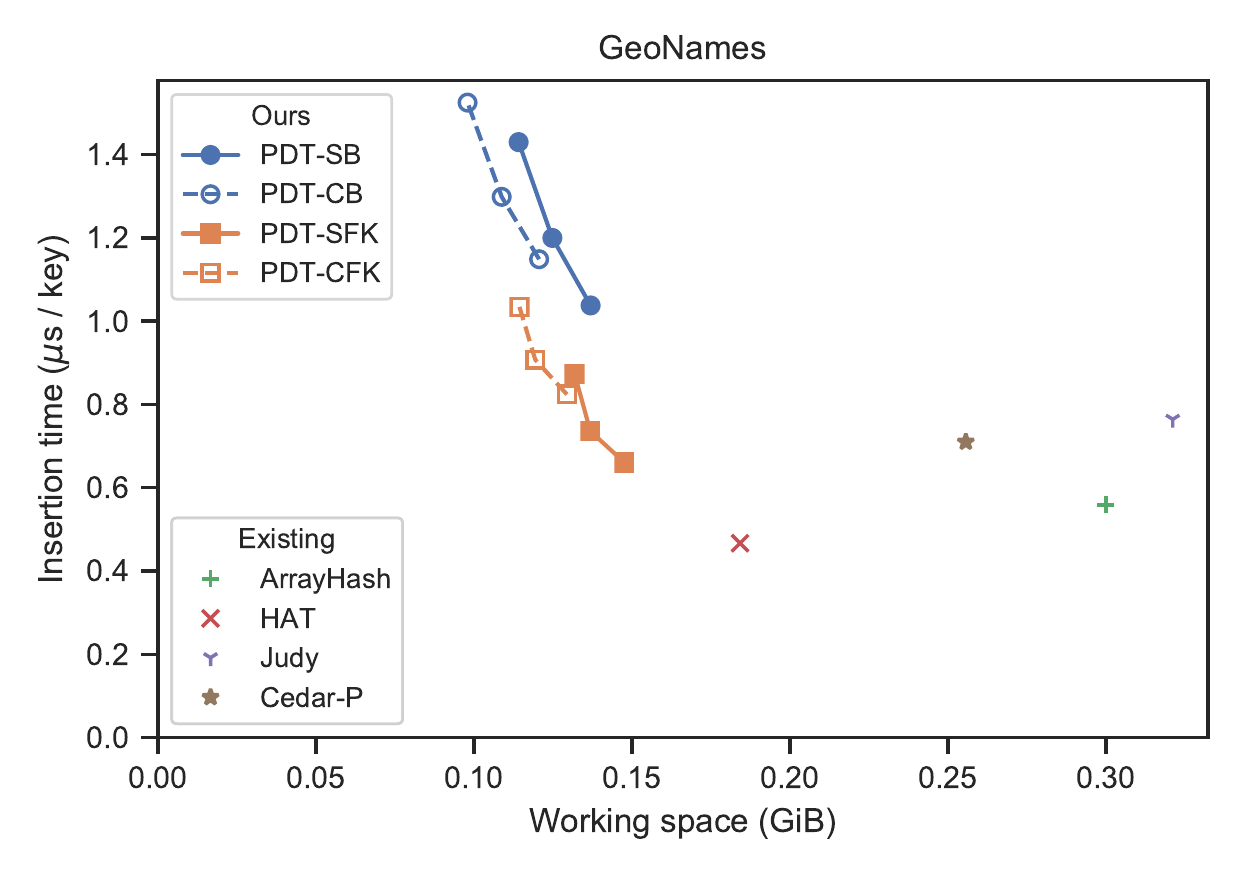}
    \includegraphics[width=\PlotWidth]{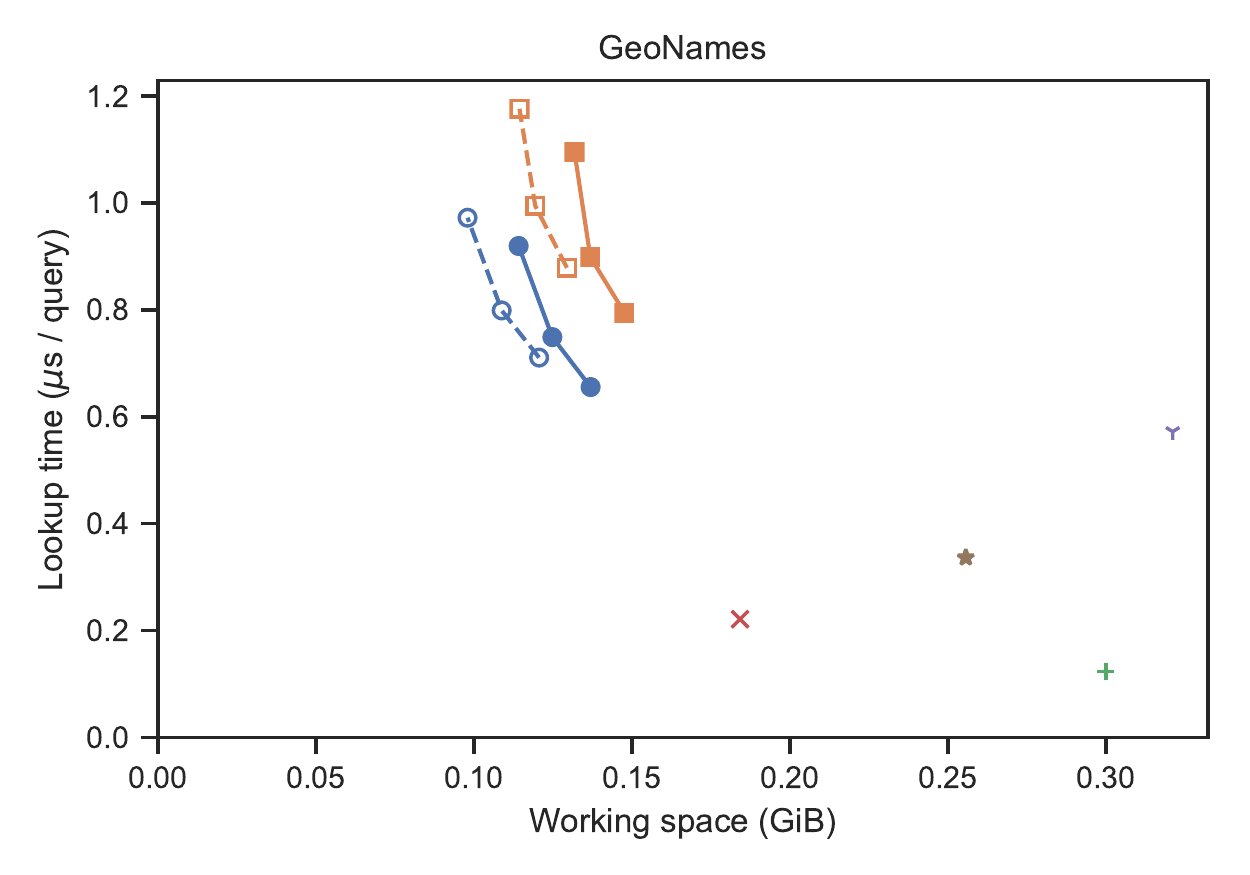}\\
    \includegraphics[width=\PlotWidth]{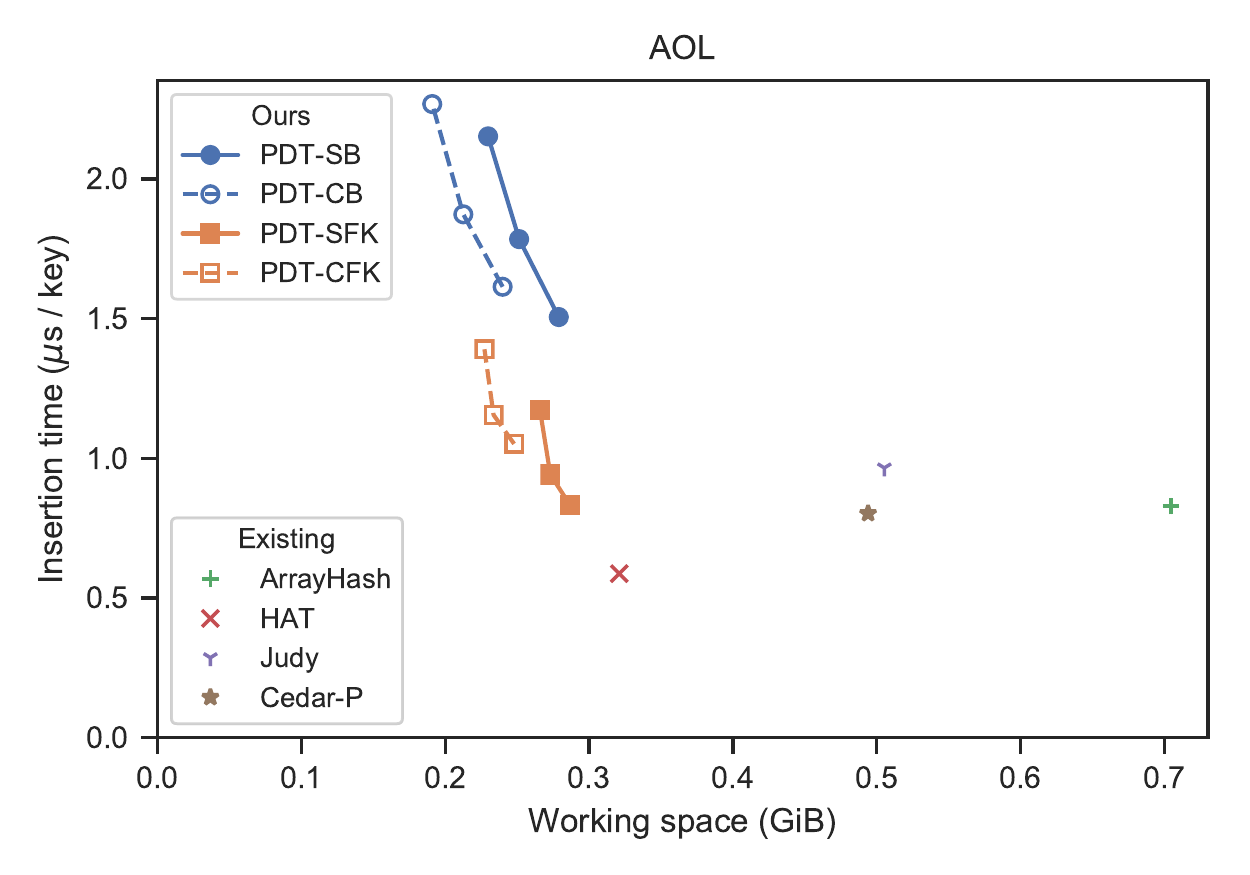}
    \includegraphics[width=\PlotWidth]{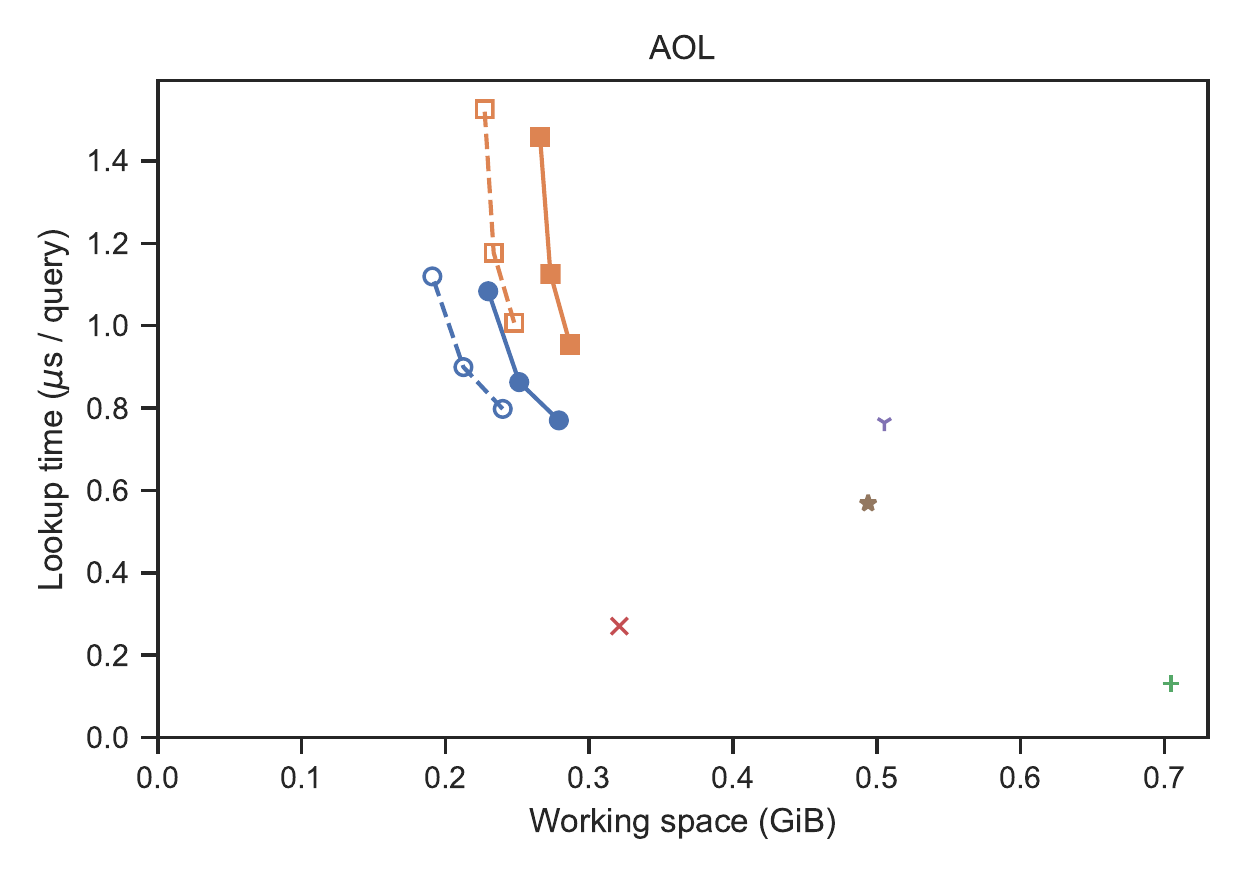}\\
    \includegraphics[width=\PlotWidth]{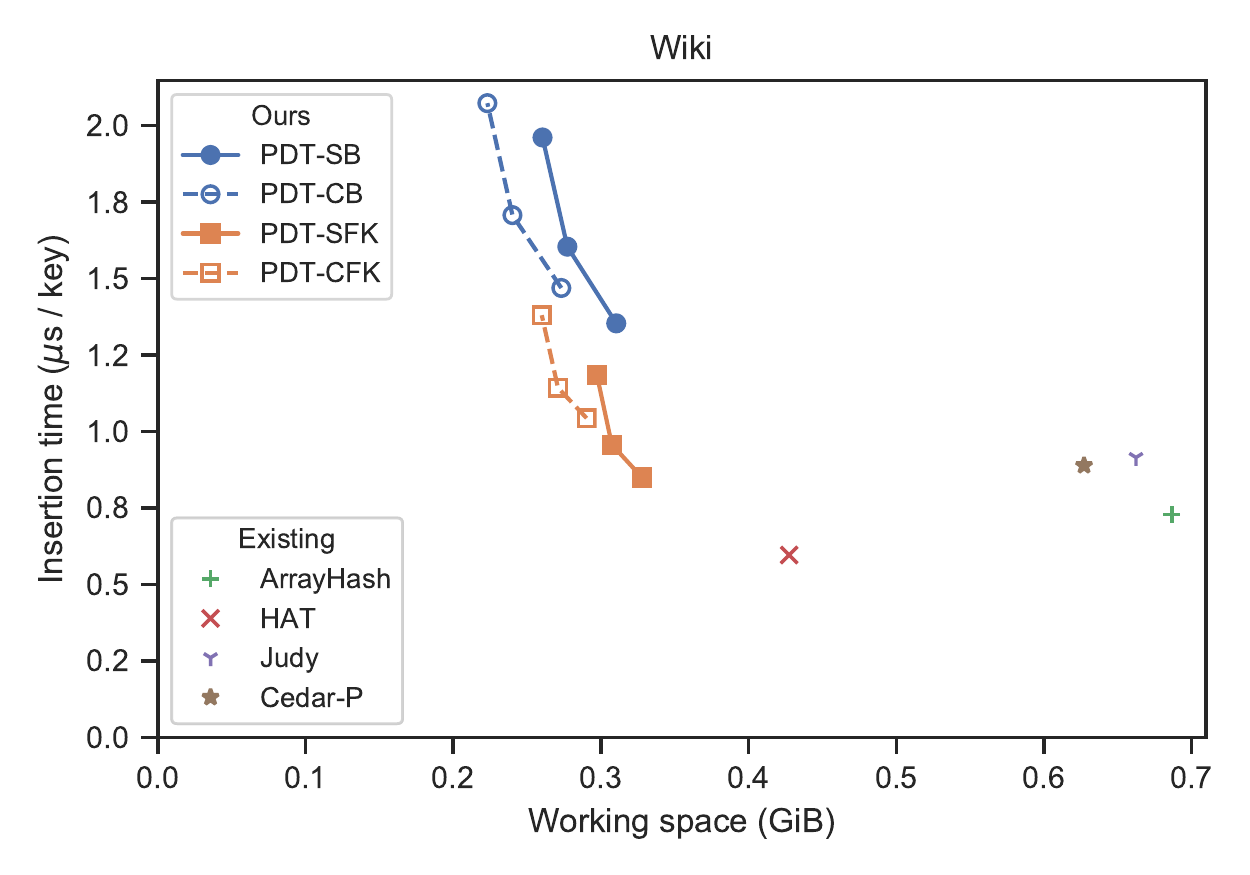}
    \includegraphics[width=\PlotWidth]{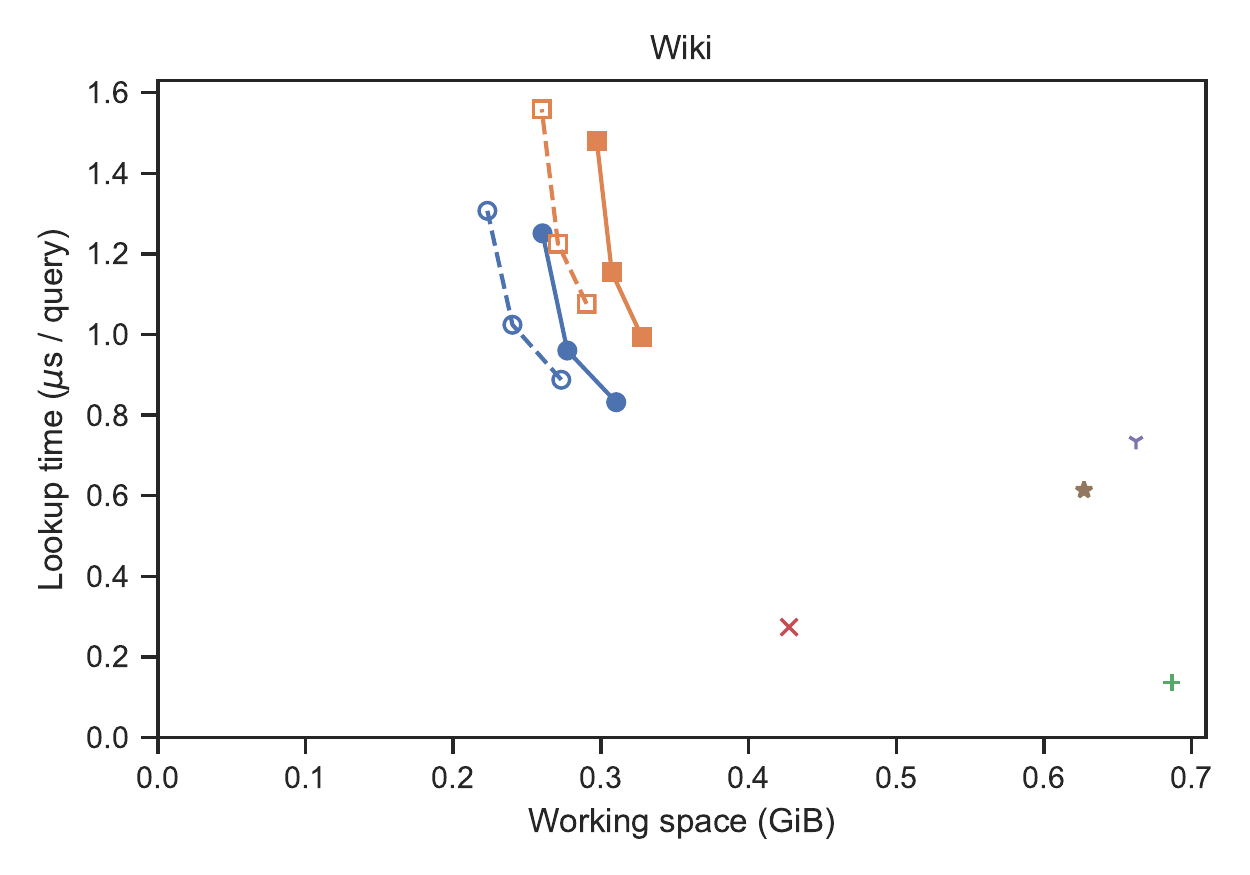}\\
    \includegraphics[width=\PlotWidth]{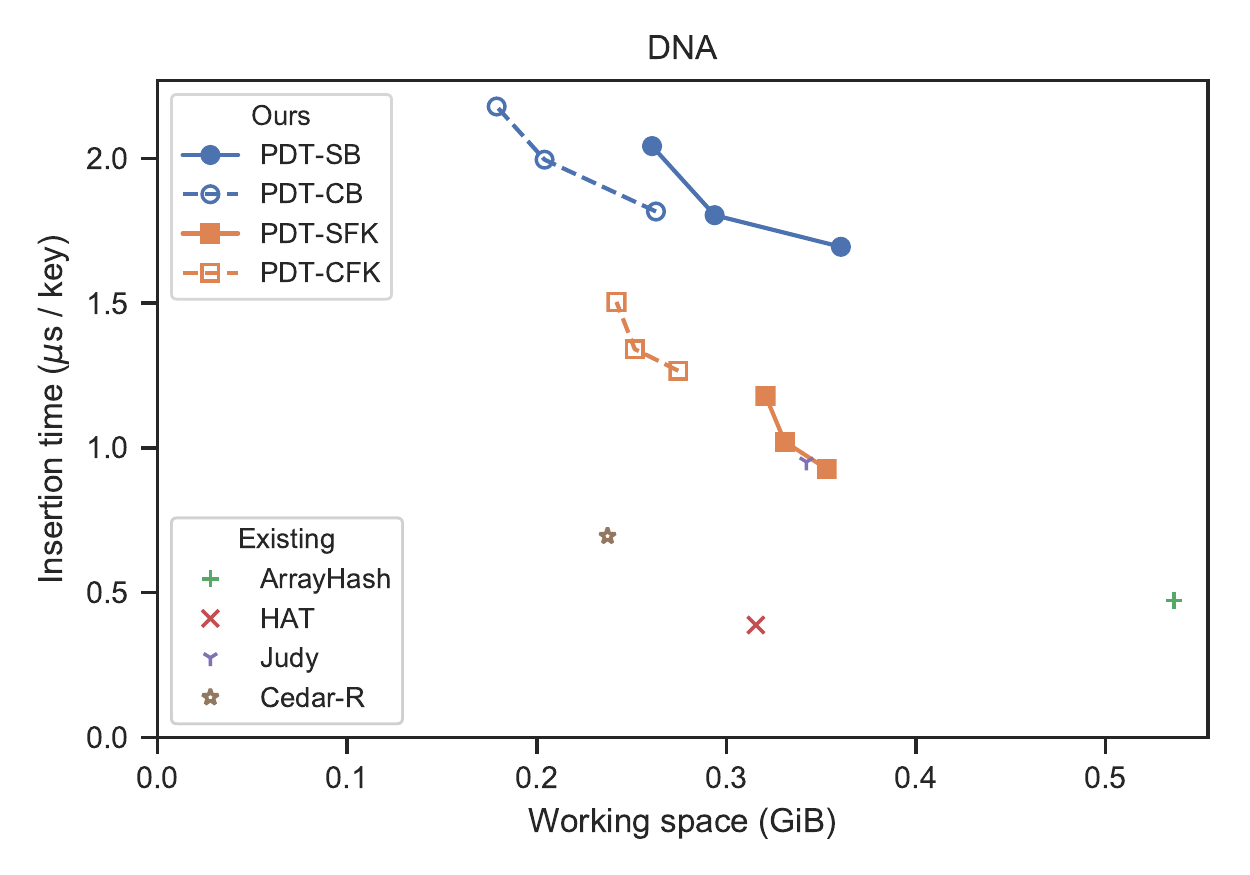}
    \includegraphics[width=\PlotWidth]{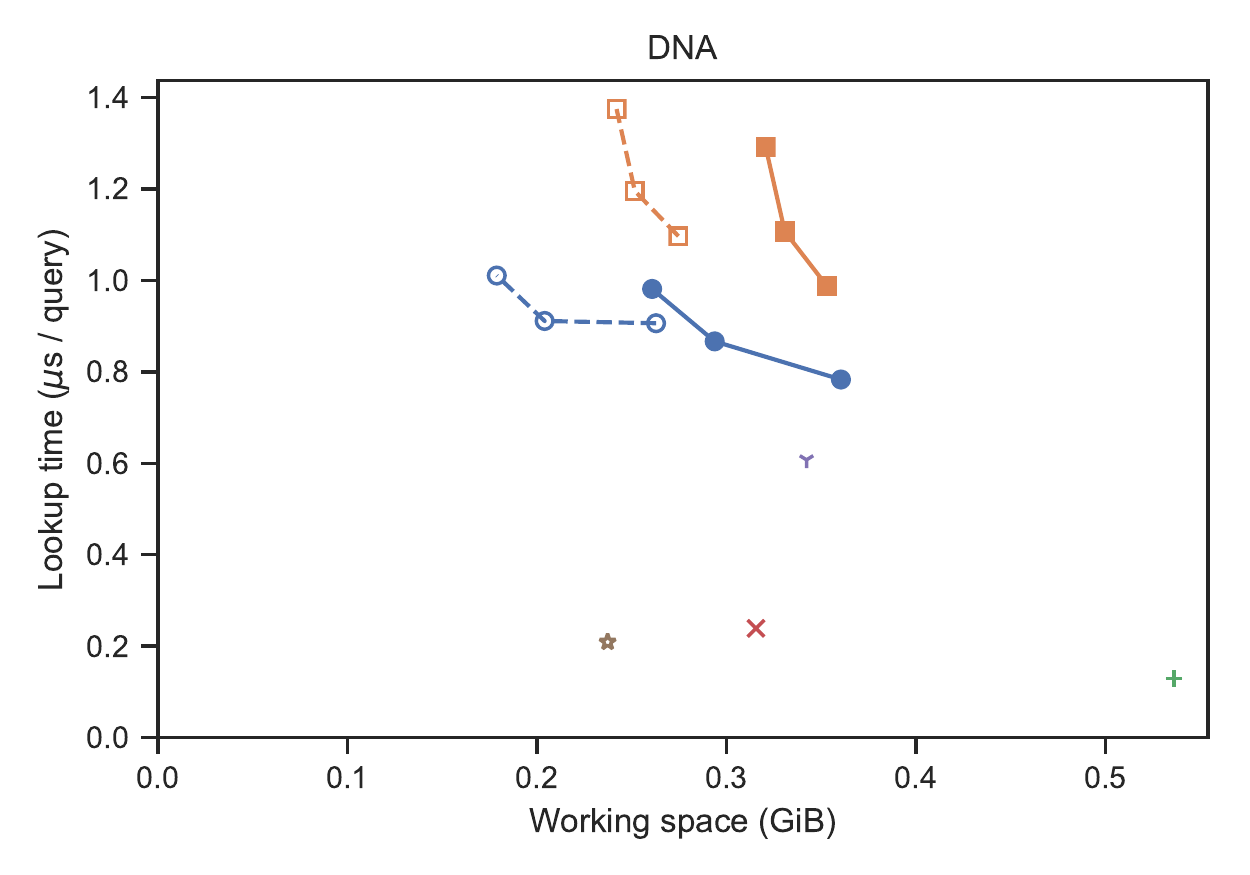}
    \caption{Experimental results for short keywords with $\ell = 8$, $16$, and $32$.}
    \label{fig:results-short}
\end{figure}

\begin{figure}[p]
    \centering
    \includegraphics[width=\PlotWidth]{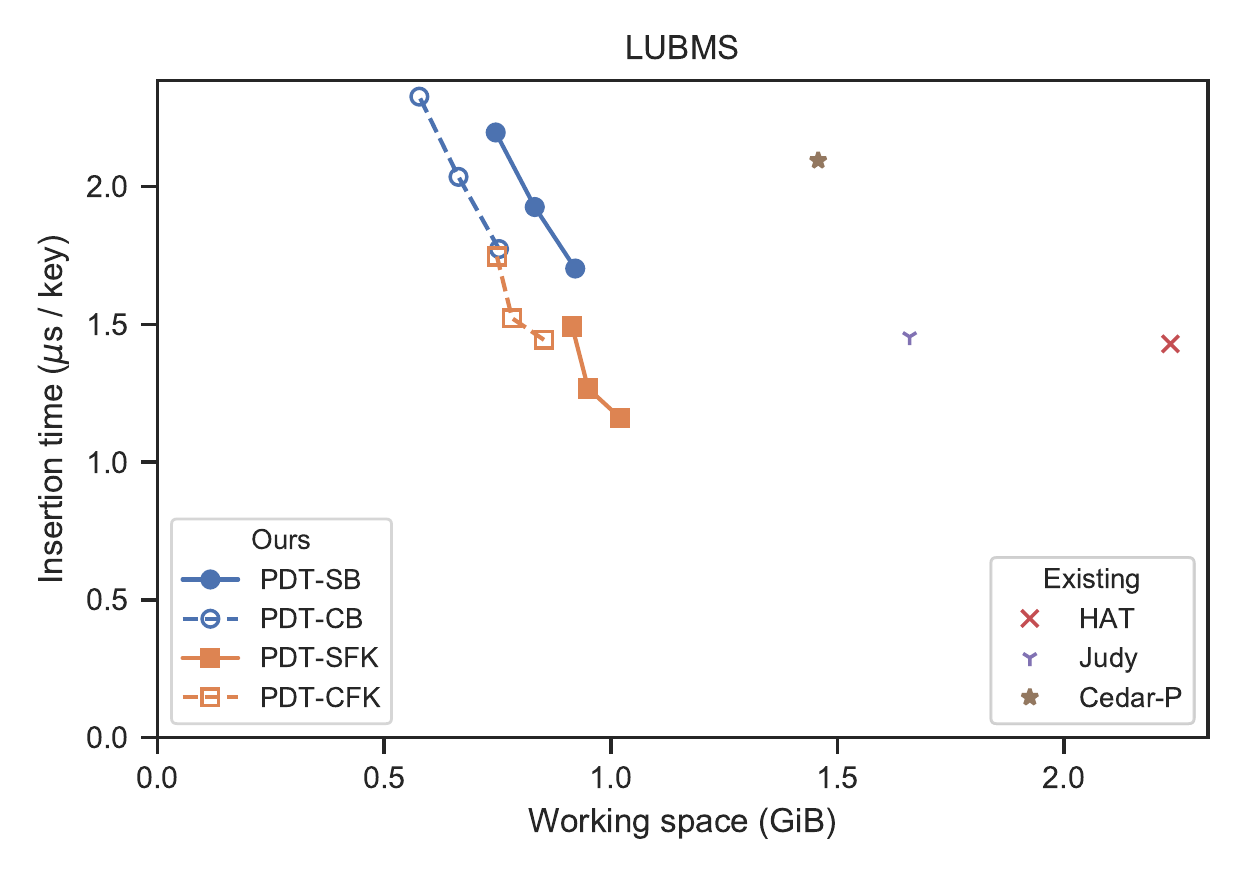}
    \includegraphics[width=\PlotWidth]{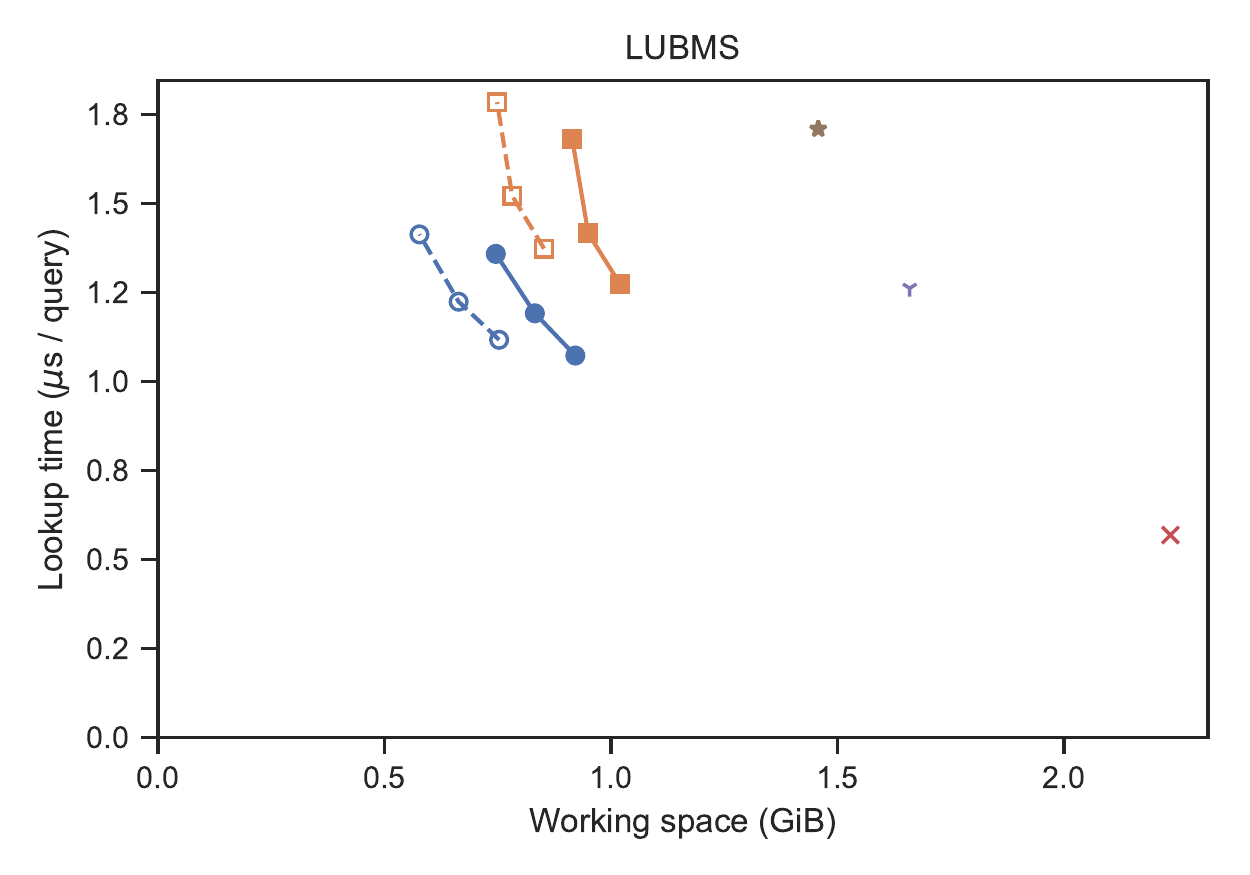}\\
    \includegraphics[width=\PlotWidth]{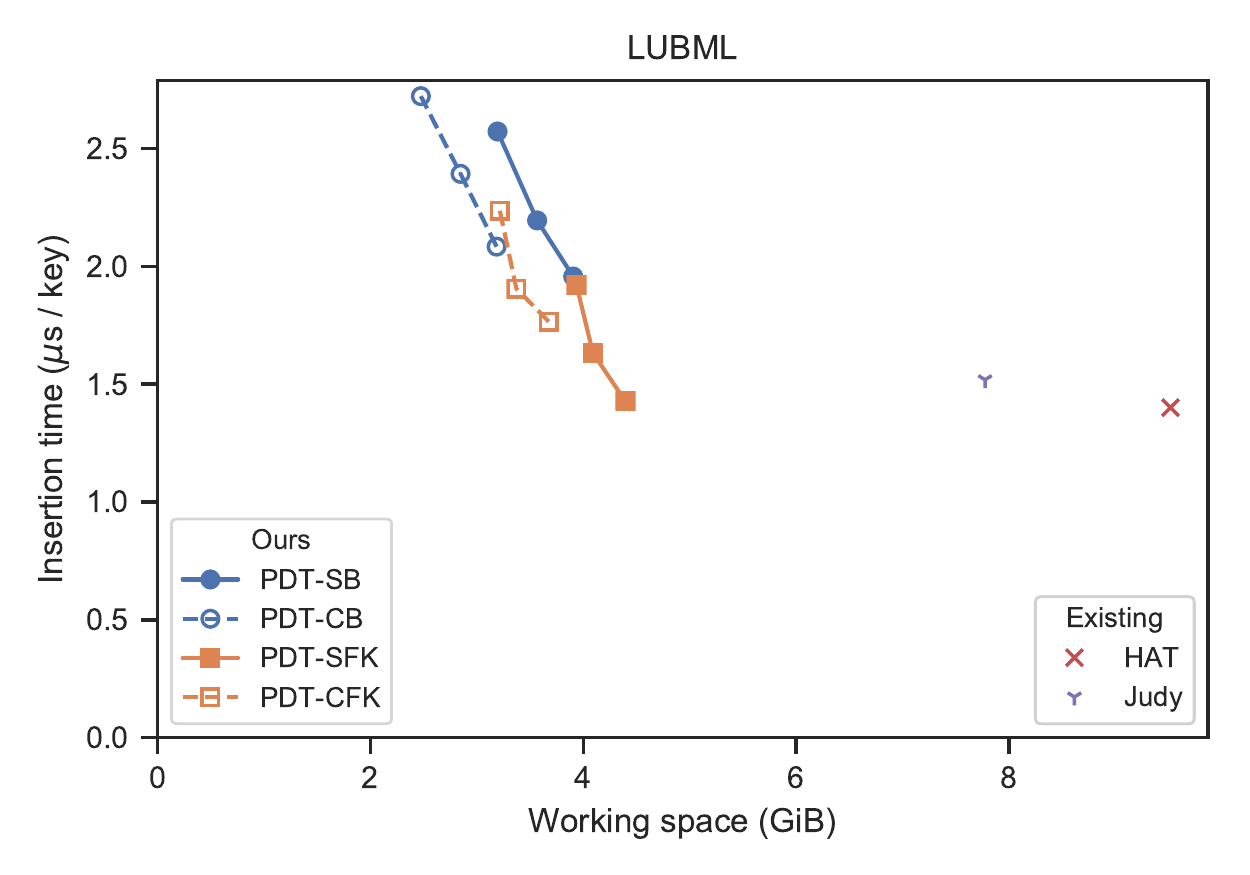}
    \includegraphics[width=\PlotWidth]{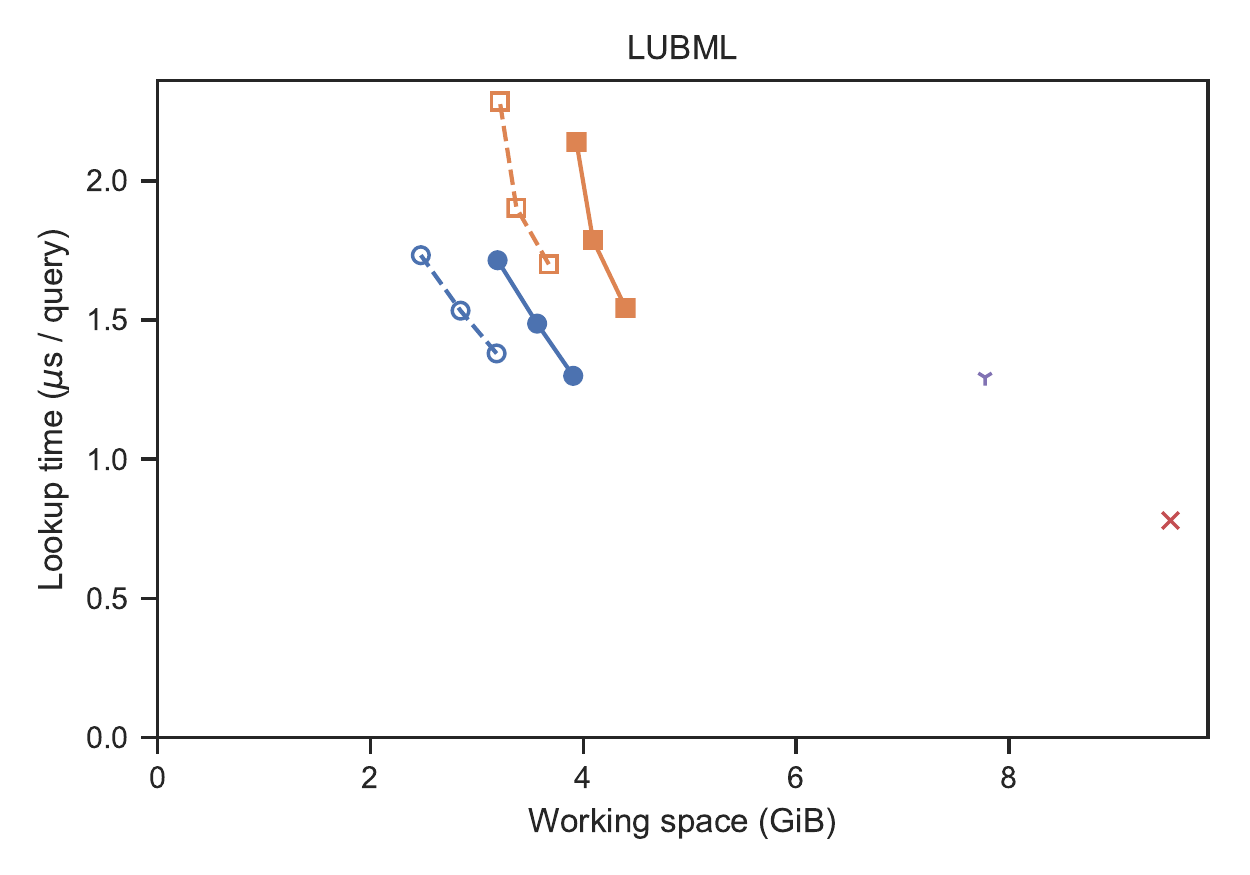}\\
    \includegraphics[width=\PlotWidth]{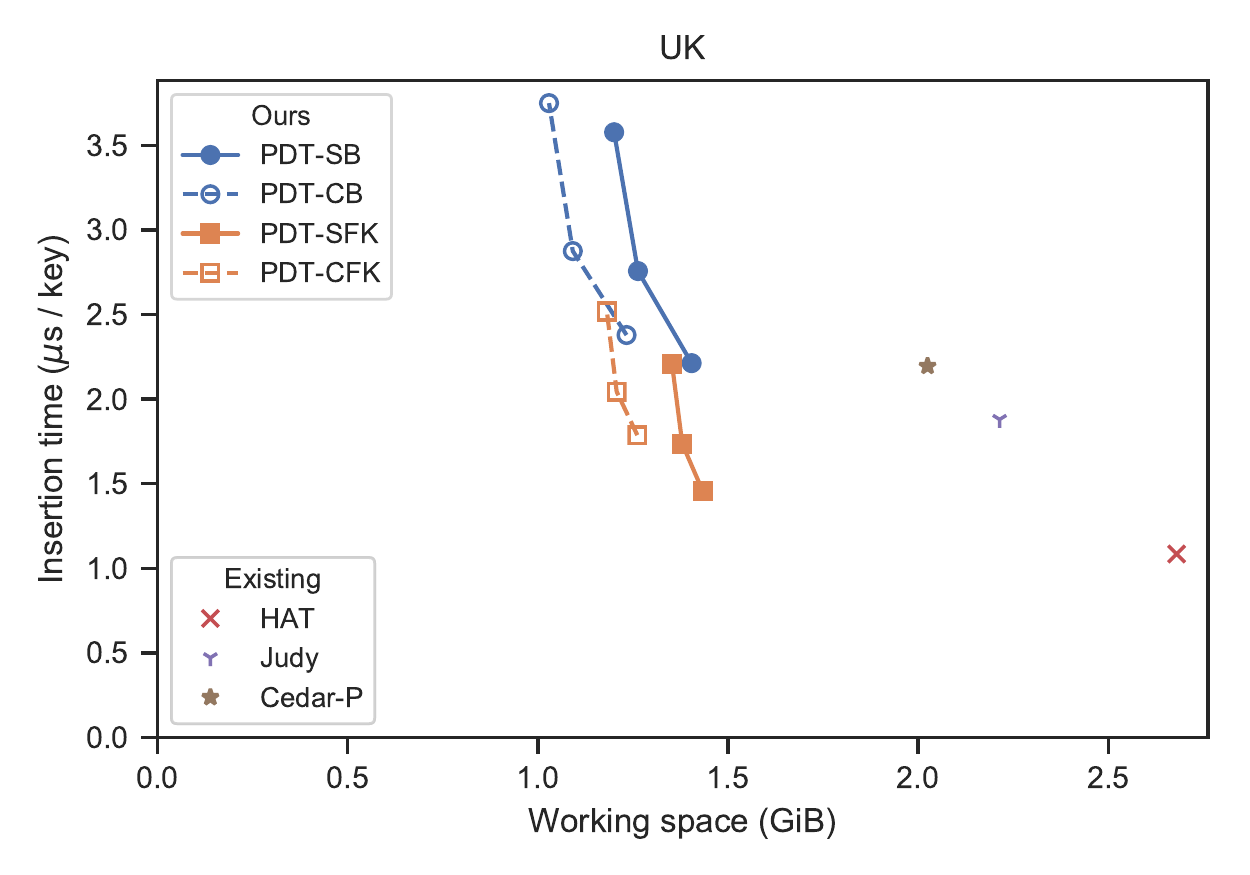}
    \includegraphics[width=\PlotWidth]{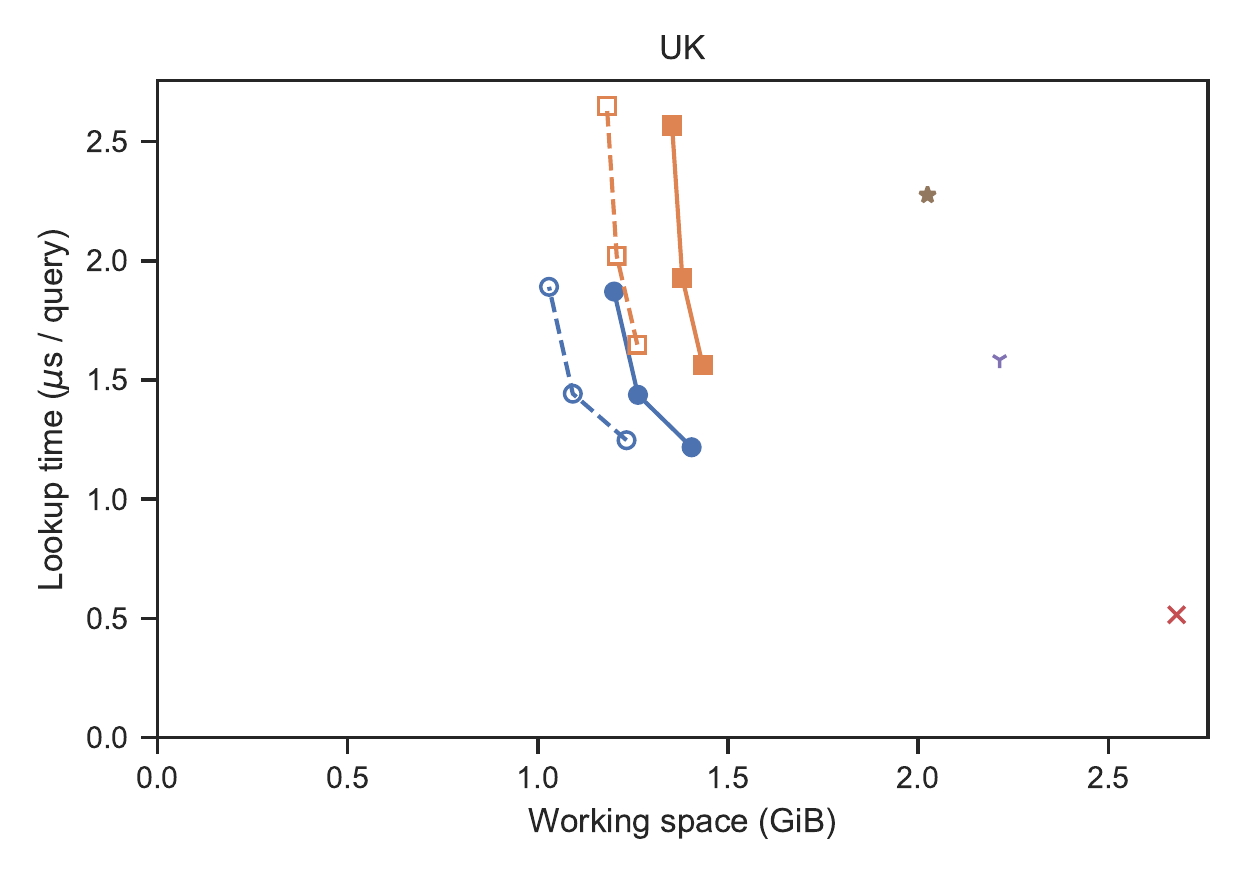}\\
    \includegraphics[width=\PlotWidth]{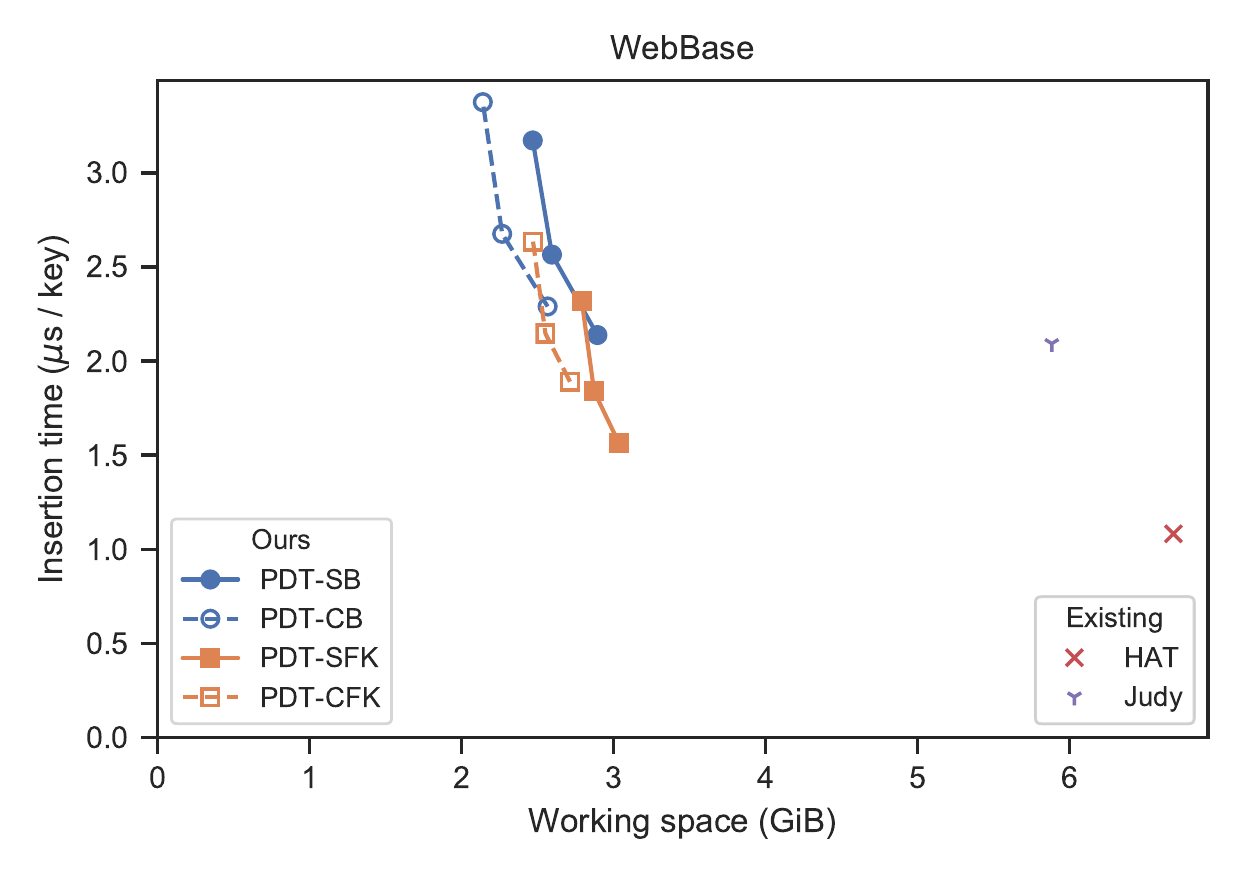}
    \includegraphics[width=\PlotWidth]{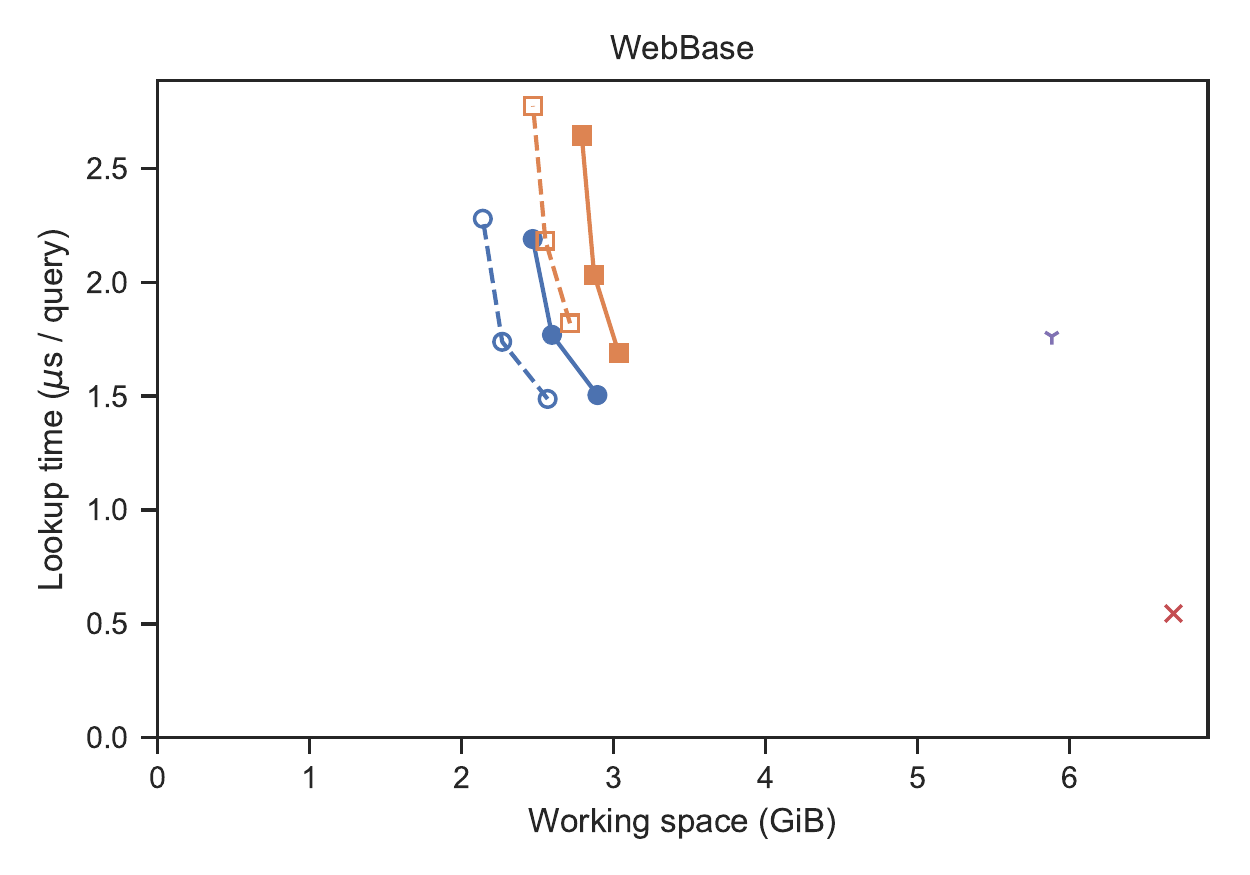}
    \caption{Experimental results for long keywords with $\ell = 8$, $16$, and $32$.}
    \label{fig:results-long}
\end{figure}

We evaluated the working space and the running times in the same manner as \sref{sect:exp:rep}.
\fref{fig:results-short} shows the experimental results for the four datasets \Geo{}, \AOL{}, \Wiki{}, and \DNA{} consisting of short keywords.
\fref{fig:results-long} shows the experimental results for the four datasets \LUBMS{}, \LUBML{}, \UK{}, and \WebBase{} consisting of long keywords.
For our methods, we only plot the results of \PCBT{}, \CCBT{}, \PCHT{} and \CCHT{}, setting $\ell$ to 8, 16, or 32.
To keep focus on the competitive contestants in the plots, we omitted some weaker instances, namely the DynPDT dictionaries with $\ell=64$ and the dictionaries with \PLM{}.
The former are too slow, while the latter take too much working space.
Only for \DNA{}, we plotted the results of \CedarR{} instead of \CedarP{} because \CedarR{} is superior on that instance.
For \LUBML{} and \WebBase{},we were not able to run our experiments with \Cedar{} because the resulting number of trie nodes becomes too large to be representable in \Cedar{} based on 32-bit pointers.
For the long keywords (\fref{fig:results-long}), we omitted the results of \ArHash{} because its working space is too large.
For example, \ArHash{} is 143\% larger than \HAT{} for \LUBMS{}.

Based on \fref{fig:results-short} showing the evaluation for short keywords,
we can state the following observations:

\begin{itemize}
    \item The DynPDT dictionaries are the smallest. \CCBT{} for $\ell = 32$ is 25--48\% smaller than the existing smallest data structures (\CedarR{} for \DNA{} and \HAT{} for the others). \CCHT{} with $\ell = 32$ is 29--39\% smaller than \HAT{} for the datasets except \DNA{}.
    \item Regarding the \Insert{} time, \HAT{} is the fastest. Except for \DNA{}, the DynPDT dictionaries based on FK-hash, \PCHT{} and \CCHT{}, are competitive to the other data structures.
    \item Regarding the \Lookup{} time, \ArHash{} is the fastest. Except for \DNA{}, the DynPDT dictionaries based on m-Bonsai, \PCBT{} and \CCBT{}, are competitive to \Judy{}.
    \item For \DNA{} consisting of short keywords, the DynPDT dictionaries are not efficient because the merits of the path decomposition applied to a trie with only short paths become negligible to the additional burden of representing the trie with two separate data structures, one for its path-decomposed trie topology and one for its node labels.
\end{itemize}

Based on \fref{fig:results-long} showing the evaluation for long keywords,
we can state the following observations:

\begin{itemize}
    \item The DynPDT dictionaries are the smallest for all the datasets. When $\ell=32$, \CCBT{} is 49--60\% smaller than \CedarP{} for \LUBMS{} and \UK{}, and is 64--68\% smaller than \Judy{} for \LUBML{} and \WebBase{}. When $\ell=32$, \CCHT{} is 42--49\% smaller than \CedarP{} for \LUBMS{} and \UK{}, and is 58--59\% smaller than \Judy{} for \LUBML{} and \WebBase{}.
    \item Regarding the \Insert{} time, \PCHT{} is competitive to the other data structures.
    \item Regarding the \Lookup{} time, \HAT{} is the fastest although its working space is large. Compared to \PCBT{} with $\ell = 8$, \HAT{} is 40--78\% faster but 48--61\% larger.
    \item In many cases, the DynPDT dictionaries outperform \Judy{} and \CedarP{}. For example, \PCHT{} with $\ell=8$ is 48\% smaller and 4--25\% faster than \Judy{} for \WebBase{}. \CCBT{} with $\ell=8$ is 48\% smaller and 15--35\% faster than \CedarP{} for \LUBMS{}.
\end{itemize}

\paragraph{Summary}

Throughout all dataset instances, DynPDT is the smallest data structure.
Especially for long keywords such as URIs, our dictionaries are space-efficient and fast thanks to the path decomposition; however, they are not efficient for extremely short keywords because the path decomposition does not work well on such instances.
In summary, DynPDT is useful for in-memory applications handling massive datasets consisting of long keywords.

For example, the RDF database system Diplodocus \cite{wylot2011diplodocus,wylot2014tripleprov} encodes every URI as an integer number through a dynamic keyword dictionary because the fixed-size integers can be handled more efficiently than the original strings having variable lengths.
Since the encoding time is a significant part of the query execution time on the Diplodocus system, Mavlyutov et al. \cite{mavlyutov2015comparison} experimentally compared a series of dynamic keyword dictionaries.
Actually, \LUBMS{} and \LUBML{} of our datasets are exactly those evaluated in \cite{mavlyutov2015comparison}.
They concluded that \HAT{} is a good data structure taking aspects like working space and time performance into account.\footnote{\Judy{} and \Cedar{} were not evaluated in \cite{mavlyutov2015comparison}.}
However, as demonstrated in our experiments, our DynPDT dictionaries can maintain the URI datasets in space up to 74\% smaller than \HAT{}, while keeping competitive insertion times.
Although DynPDT's slow lookup time is a drawback compared to \HAT{}, maintaining massive RDF database systems in main-memory is essential, and we believe that DynPDT's high memory efficiency will contribute to the future of Semantic Web applications.

\section{Conclusion}
\label{sect:con}

We presented a novel data structure for dynamic keyword dictionaries --- called DynPDT --- which is applicable to scalable string data processing.
For that, we applied path decomposition and utilized the recent hash-based trie representations m-Bonsai and FK-hash.
We demonstrated with experiments on real-world massive datasets that the memory footprint of DynPDT is the smallest within a careful selection of efficient dynamic keyword dictionaries. 
It is especially efficient for long keywords due to the path decomposition approach.

Our results pave new ways for major improvements in various existing systems because the dynamic keyword dictionary problem is a common task in applications such as vocabulary accumulation for inverted-index construction \cite{heinz2002burst}, RDF database systems \cite{wylot2011diplodocus,wylot2014tripleprov}, in-memory OLTP (online transaction processing) database systems \cite{leis2013adaptive}, Web crawlers \cite{ueda2013parallel}, and search engines \cite{software:groonga,busch2012earlybird}.
DynPDT can contribute to those systems especially by reducing their memory requirements.
Although we have put the focus on the keyword dictionary problem in this paper, DynPDT as a general data structure is of independent interest, being useful for applications handling  dynamic tries.
An interesting application is the LZD compression~\cite{goto2015lzd,badkobeh2017two}, a variation of the LZ78 compression \cite{ziv1978compression}.
Since the LZD algorithm maintains long factors (or strings) in a dynamic trie, we are confident that the incremental path decomposition on such a trie will have performance benefits.

Our future plans for DynPDT are as follows.

\begin{itemize}
\item The \emph{burst trie} developed by Heinz et al. \cite{heinz2002burst} maintains sparse subtries in a trie in dynamic containers of strings by collapsing the subtries. DynPDT would be suited as an alternative container representation to enhance the memory efficiency of the burst trie.

\item In our experiments, we implemented the second data structure of the displacement array $D_2$ through the CHT by Cleary \cite{cleary1984compact}, following the original m-Bonsai approach \cite{poyias2018mbonsai}.
Recently, K{\"{o}}ppl et al. \cite{koeppl2020fast} developed space-efficient hash tables with separate chaining and compact hashing.
Although the CHT needs additional displacement information (i.e., two bit arrays), his hash tables do not need such additional information.
We expect that his hash tables are suitable representations of $D_2$.

\end{itemize}

\begin{acks}
We thank Kazuya Tsuruta for kindly providing us the implementations used in \cite{tsuruta2020ctrie}.
We thank the anonymous reviewers for their helpful comments.
A part of this work was supported by JSPS KAKENHI Grant Numbers 17J07555 and JP18F18120.
\end{acks}


\appendix

\section{Experimental Results}
\label{apdx:exp}
Within the same setting as in \sref{sect:exp:exist}, we present an extended evaluation including the following contestants:

\begin{itemize}
    \item \STL{} is the hash table \texttt{std::unordered\_map} of the C++ standard library.
    \item \GDense{} is the hash table implementation \texttt{google::dense\_hash\_map} of Google~\cite{software:sparsehash}.
    \item \GSparse{} is Gregory Popovitch's space-efficient hash container implementation derived from Google's sparse hash table \cite{software:sparsepp}.
    \item \Hopscotch{} is Tessil's  hash table implementation using hopscotch hashing \cite{software:hopscotch-map}.
    \item \Robin{} is Tessil's hash table implementation using robin hood hashing \cite{software:robin-map}.
    \item \ART{} is Armon Dadgar's implementation \cite{software:libart} of the adaptive radix tree \cite{leis2013adaptive}.
\end{itemize}

Further, we include the following implementations, which are also used and studied in the experimental section of \cite{tsuruta2019dynamic}:

\begin{itemize}
    \item \PCTB{} is a packed compact trie using bit parallelism \cite{takagi2016packed}.
    \item \PCTH{} is a packed compact trie using additionally \STL{} as a dictionary in each micro trie \cite{takagi2016packed}.
    \item \ZFT{} is Tsuruta's C++ implementation of 
    the z-fast trie \cite{belazzougui2010dynamic}.
    \item \CTrie{} is a trie~\cite{tsuruta2019dynamic} combining aspects of the z-fast trie with the packed compact trie.
\end{itemize}

\tref{tab:appex-short} shows the results for the datasets consisting of short keywords (i.e., \Geo{}, \AOL{}, \Wiki{} and \DNA{}).
\tref{tab:appex-long} shows the results for the datasets consisting of long keywords (i.e., \LUBMS{}, \LUBML{}, \UK{} and \WebBase{}).
In these tables, \Space{} is the working space in GiB, \InsertTime{} is the average insertion time in microseconds, and \LookupTime{} is the average lookup time in microseconds.
For \CLM{} of DynPDT, the results with $\ell = 16$ are shown.
Concerning \PCTB{}, \PCTH{}, \ZFT{} and \CTrie{}, we could not obtain some results for large datasets because the resulting trie was too large to fit into RAM\@.

\begin{table}[p]
\footnotesize
\centering
\caption{Experimental results for short keywords.}
\label{tab:appex-short}
\subfloat[\Geo{}]{
    \label{tab:geo}
    \begin{tabular}{lrrr}
\toprule
 & \Space{} & \InsertTime{} & \LookupTime{} \\
\midrule
\PPBT{} & 0.32 & 0.59 & 0.53 \\
\PCBT{} & 0.12 & 1.20 & 0.75 \\
\CCBT{} & 0.11 & 1.30 & 0.80 \\
\PPHT{} & 0.33 & 0.63 & 0.71 \\
\PCHT{} & 0.14 & 0.74 & 0.90 \\
\CCHT{} & 0.12 & 0.91 & 0.99 \\
\midrule
\STL{} & 0.58 & 0.44 & 0.24 \\
\GDense{} & 0.73 & 0.37 & 0.12 \\
\GSparse{} & 0.42 & 0.58 & 0.15 \\
\Hopscotch{} & 0.84 & 0.40 & 0.10 \\
\Robin{} & 0.84 & 0.31 & 0.10 \\
\ArHash{} & 0.30 & 0.56 & 0.12 \\
\HAT{} & 0.18 & 0.47 & 0.22 \\
\Judy{} & 0.32 & 0.76 & 0.57 \\
\ART{} & 0.59 & 0.85 & 0.56 \\
\CedarR{} & 0.47 & 0.68 & 0.39 \\
\CedarP{} & 0.26 & 0.71 & 0.34 \\
\midrule
\PCTB{} & 2.96 & 9.12 & 10.43 \\
\PCTH{} & 5.13 & 7.74 & 5.49 \\
\ZFT{} & 1.13 & 3.25 & 2.42 \\
\CTrie{} & 1.34 & 2.58 & 0.79 \\
\bottomrule
\end{tabular}
}
\subfloat[\AOL{}]{
    \label{tab:aol}
    \begin{tabular}{lrrr}
\toprule
 & \Space{} & \InsertTime{} & \LookupTime{} \\
\midrule
\PPBT{} & 0.52 & 1.01 & 0.65 \\
\PCBT{} & 0.25 & 1.78 & 0.86 \\
\CCBT{} & 0.21 & 1.87 & 0.90 \\
\PPHT{} & 0.52 & 0.80 & 0.80 \\
\PCHT{} & 0.27 & 0.94 & 1.13 \\
\CCHT{} & 0.23 & 1.16 & 1.18 \\
\midrule
\STL{} & 1.01 & 0.52 & 0.26 \\
\GDense{} & 1.72 & 0.67 & 0.15 \\
\GSparse{} & 0.77 & 0.76 & 0.18 \\
\Hopscotch{} & 1.04 & 0.51 & 0.12 \\
\Robin{} & 1.79 & 0.47 & 0.12 \\
\ArHash{} & 0.70 & 0.83 & 0.13 \\
\HAT{} & 0.32 & 0.59 & 0.27 \\
\Judy{} & 0.51 & 0.97 & 0.76 \\
\ART{} & 0.91 & 0.96 & 0.77 \\
\CedarR{} & 1.07 & 0.87 & 0.61 \\
\CedarP{} & 0.49 & 0.80 & 0.57 \\
\midrule
\PCTB{} & 4.07 & 12.42 & 14.34 \\
\PCTH{} & 7.66 & 10.26 & 6.99 \\
\ZFT{} & 1.82 & 3.84 & 2.57 \\
\CTrie{} & 2.12 & 3.01 & 1.10 \\
\bottomrule
\end{tabular}
}\\
\subfloat[\Wiki{}]{
    \label{tab:wiki}
    \begin{tabular}{lrrr}
\toprule
 & \Space{} & \InsertTime{} & \LookupTime{} \\
\midrule
\PPBT{} & 0.64 & 0.98 & 0.68 \\
\PCBT{} & 0.28 & 1.60 & 0.96 \\
\CCBT{} & 0.24 & 1.71 & 1.02 \\
\PPHT{} & 0.67 & 0.79 & 0.86 \\
\PCHT{} & 0.31 & 0.96 & 1.15 \\
\CCHT{} & 0.27 & 1.14 & 1.22 \\
\midrule
\STL{} & 1.29 & 0.50 & 0.27 \\
\GDense{} & 1.64 & 0.54 & 0.14 \\
\GSparse{} & 0.97 & 0.69 & 0.18 \\
\Hopscotch{} & 1.08 & 0.42 & 0.13 \\
\Robin{} & 1.83 & 0.41 & 0.12 \\
\ArHash{} & 0.69 & 0.73 & 0.14 \\
\HAT{} & 0.43 & 0.60 & 0.27 \\
\Judy{} & 0.66 & 0.92 & 0.74 \\
\ART{} & 1.23 & 1.00 & 0.73 \\
\CedarR{} & 1.19 & 0.89 & 0.59 \\
\CedarP{} & 0.63 & 0.89 & 0.61 \\
\midrule
\PCTB{} & 5.67 & 11.85 & 13.79 \\
\PCTH{} & 10.11 & 9.48 & 7.09 \\
\ZFT{} & 2.24 & 3.56 & 2.64 \\
\CTrie{} & 2.92 & 3.01 & 1.09 \\
\bottomrule
\end{tabular}
}
\subfloat[\DNA{}]{
    \label{tab:dna}
    \begin{tabular}{lrrr}
\toprule
 & \Space{} & \InsertTime{} & \LookupTime{} \\
\midrule
\PPBT{} & 0.84 & 1.18 & 0.65 \\
\PCBT{} & 0.29 & 1.80 & 0.87 \\
\CCBT{} & 0.20 & 2.00 & 0.91 \\
\PPHT{} & 0.80 & 0.85 & 0.88 \\
\PCHT{} & 0.33 & 1.02 & 1.11 \\
\CCHT{} & 0.25 & 1.34 & 1.20 \\
\midrule
\STL{} & 1.02 & 0.91 & 0.34 \\
\GDense{} & 1.25 & 0.24 & 0.09 \\
\GSparse{} & 0.67 & 0.50 & 0.13 \\
\Hopscotch{} & 1.50 & 0.27 & 0.07 \\
\Robin{} & 1.50 & 0.26 & 0.08 \\
\ArHash{} & 0.54 & 0.47 & 0.13 \\
\HAT{} & 0.32 & 0.39 & 0.24 \\
\Judy{} & 0.34 & 0.95 & 0.61 \\
\ART{} & 1.01 & 0.65 & 0.63 \\
\CedarR{} & 0.24 & 0.70 & 0.21 \\
\CedarP{} & 0.31 & 0.67 & 0.24 \\
\midrule
\PCTB{} & 7.05 & 8.44 & 9.91 \\
\PCTH{} & 8.45 & 5.44 & 6.86 \\
\ZFT{} & 2.57 & 3.30 & 2.88 \\
\CTrie{} & 2.50 & 2.55 & 0.78 \\
\bottomrule
\end{tabular}
}
\end{table}

\begin{table}[p]
\footnotesize
\centering
\caption{Experimental results for long keywords.}
\label{tab:appex-long}
\subfloat[\LUBMS{}]{
    \label{tab:lubms}
    \begin{tabular}{lrrr}
\toprule
 & \Space{} & \InsertTime{} & \LookupTime{} \\
\midrule
\PPBT{} & 2.37 & 1.62 & 1.10 \\
\PCBT{} & 0.83 & 1.93 & 1.19 \\
\CCBT{} & 0.66 & 2.04 & 1.22 \\
\PPHT{} & 2.46 & 1.09 & 1.14 \\
\PCHT{} & 0.95 & 1.27 & 1.42 \\
\CCHT{} & 0.78 & 1.52 & 1.52 \\
\midrule
\STL{} & 7.47 & 0.61 & 0.51 \\
\GDense{} & 9.93 & 0.89 & 0.28 \\
\GSparse{} & 6.22 & 0.83 & 0.39 \\
\Hopscotch{} & 6.87 & 0.70 & 0.27 \\
\Robin{} & 9.87 & 0.61 & 0.26 \\
\ArHash{} & 5.44 & 0.98 & 0.30 \\
\HAT{} & 2.23 & 1.43 & 0.57 \\
\Judy{} & 1.66 & 1.45 & 1.26 \\
\ART{} & 5.83 & 0.91 & 0.77 \\
\CedarR{} & 1.97 & 1.79 & 1.66 \\
\CedarP{} & 1.46 & 2.10 & 1.71 \\
\midrule
\PCTB{} & n/a & n/a & n/a \\
\PCTH{} & n/a & n/a & n/a \\
\ZFT{} & 9.27 & 6.33 & 5.65 \\
\CTrie{} & 8.13 & 4.25 & 2.43 \\
\bottomrule
\end{tabular}
}
\subfloat[\LUBML{}]{
    \label{tab:lubml}
    \begin{tabular}{lrrr}
\toprule
 & \Space{} & \InsertTime{} & \LookupTime{} \\
\midrule
\PPBT{} & 10.1 & 1.43 & 1.00 \\
\PCBT{} & 3.6 & 2.20 & 1.49 \\
\CCBT{} & 2.8 & 2.39 & 1.53 \\
\PPHT{} & 10.7 & 1.32 & 1.39 \\
\PCHT{} & 4.1 & 1.63 & 1.79 \\
\CCHT{} & 3.4 & 1.91 & 1.90 \\
\midrule
\STL{} & 32.9 & 0.67 & 0.59 \\
\GDense{} & 40.1 & 0.99 & 0.32 \\
\GSparse{} & 27.3 & 0.89 & 0.44 \\
\Hopscotch{} & 41.2 & 1.01 & 0.27 \\
\Robin{} & 41.2 & 0.69 & 0.28 \\
\ArHash{} & 21.9 & 1.06 & 0.32 \\
\HAT{} & 9.5 & 1.40 & 0.78 \\
\Judy{} & 7.8 & 1.52 & 1.29 \\
\ART{} & 25.8 & 1.05 & 0.93 \\
\CedarR{} & n/a & n/a & n/a \\
\CedarP{} & n/a & n/a & n/a \\
\midrule
\PCTB{} & n/a & n/a & n/a \\
\PCTH{} & n/a & n/a & n/a \\
\ZFT{} & n/a & n/a & n/a \\
\CTrie{} & n/a & n/a & n/a \\
\bottomrule
\end{tabular}
}\\
\subfloat[\UK{}]{
    \label{tab:uk}
    \begin{tabular}{lrrr}
\toprule
 & \Space{} & \InsertTime{} & \LookupTime{} \\
\midrule
\PPBT{} & 2.32 & 1.45 & 0.94 \\
\PCBT{} & 1.26 & 2.76 & 1.44 \\
\CCBT{} & 1.09 & 2.87 & 1.44 \\
\PPHT{} & 2.32 & 1.27 & 1.24 \\
\PCHT{} & 1.38 & 1.74 & 1.93 \\
\CCHT{} & 1.21 & 2.04 & 2.02 \\
\midrule
\STL{} & 6.05 & 0.67 & 0.50 \\
\GDense{} & 10.50 & 1.09 & 0.27 \\
\GSparse{} & 5.06 & 0.96 & 0.37 \\
\Hopscotch{} & 6.23 & 0.75 & 0.25 \\
\Robin{} & 9.23 & 0.63 & 0.25 \\
\ArHash{} & 5.91 & 1.16 & 0.28 \\
\HAT{} & 2.68 & 1.08 & 0.51 \\
\Judy{} & 2.21 & 1.88 & 1.59 \\
\ART{} & 5.17 & 1.64 & 1.19 \\
\CedarR{} & 7.37 & 2.24 & 2.30 \\
\CedarP{} & 2.02 & 2.20 & 2.28 \\
\midrule
\PCTB{} & 18.05 & 25.92 & 33.49 \\
\PCTH{} & n/a & n/a & n/a \\
\ZFT{} & 7.53 & 6.20 & 5.03 \\
\CTrie{} & 8.17 & 4.75 & 2.86 \\
\bottomrule
\end{tabular}
}
\subfloat[\WebBase{}]{
    \label{tab:webbase}
    \begin{tabular}{lrrr}
\toprule
 & \Space{} & \InsertTime{} & \LookupTime{} \\
\midrule
\PPBT{} & 5.4 & 1.66 & 1.19 \\
\PCBT{} & 2.6 & 2.57 & 1.77 \\
\CCBT{} & 2.3 & 2.68 & 1.74 \\
\PPHT{} & 5.7 & 1.36 & 1.42 \\
\PCHT{} & 2.9 & 1.84 & 2.03 \\
\CCHT{} & 2.6 & 2.15 & 2.18 \\
\midrule
\STL{} & 16.3 & 0.64 & 0.55 \\
\GDense{} & 19.5 & 0.82 & 0.28 \\
\GSparse{} & 13.5 & 0.83 & 0.43 \\
\Hopscotch{} & 20.3 & 0.93 & 0.24 \\
\Robin{} & 20.3 & 0.64 & 0.26 \\
\ArHash{} & 10.4 & 0.98 & 0.29 \\
\HAT{} & 6.7 & 1.08 & 0.55 \\
\Judy{} & 5.9 & 2.09 & 1.76 \\
\ART{} & 14.0 & 1.76 & 1.45 \\
\CedarR{} & n/a & n/a & n/a \\
\CedarP{} & n/a & n/a & n/a \\
\midrule
\PCTB{} & n/a & n/a & n/a \\
\PCTH{} & n/a & n/a & n/a \\
\ZFT{} & 19.6 & 5.77 & 5.06 \\
\CTrie{} & 23.5 & 5.12 & 3.12 \\
\bottomrule
\end{tabular}
}
\end{table}

\end{document}